\documentclass[12pt]{article}
\usepackage{amsfonts}
\usepackage{amsmath}
\usepackage{amssymb}
\usepackage{amsthm}

\usepackage[utf8]{inputenc}
\usepackage[margin=2cm]{geometry}
\usepackage{geometry}
\usepackage{hyperref}
\usepackage{graphicx}
\usepackage{amsmath}
\usepackage{cite}
\usepackage{hyperref}
\usepackage{tensor}
\usepackage{cases}
\usepackage{appendix}
\usepackage{multirow}
\usepackage{booktabs}
\usepackage{color}
\usepackage[font=small]{subfig}
\usepackage[skip=8pt,labelfont={bf,sc},textfont=small]{caption}

\def\[#1\]{\begin{align}#1\end{align}}
\newcommand{\imineq}[2]{\vcenter{\hbox{\includegraphics[height=#2ex]{#1}}}}
\def \nn {\nonumber}
\def \a {\alpha}
\def \b {\beta}
\def \g {\gamma}

\def \d {\delta}

\def \e {\epsilon}

\def \pd{\partial}

\def \ra{\rangle}
\def \la{\langle}

\def \mo{\mathcal{O}}

\def \Tr{\text{tr}}
\def \tr{\text{tr}}
\def \T{\mathcal{T}}
\def \N{\mathcal{N}}

\def \tpsi{\tilde{\psi}}
\def \S{\mathcal{S}}
\def \L{\mathcal{L}}

\begin{document}
\begin{titlepage}
%\begin{flushright}
%TIT/HEP-6XX \\
%mm,  2016
%\end{flushright}
\vspace{0.5cm}
\begin{center}
%{\Large \bf {Quantum entanglement of superposition states and fusion rules in RCFT local operator quenches, Entanglement Dynamics in RCFT: Fusion Rules, Operator Product Expansion, and Quasiparticle Propagation}}
{\Large \bf {Entanglement and Pseudo Entanglement Dynamics versus Fusion in CFT}}
\lineskip .75em
\vskip 2.5cm
{\large Song He$^{a,b,}$\footnote{hesong@jlu.edu.cn},  Yu-Xuan Zhang$^{a,}$\footnote{yuxuanz20@mails.jlu.edu.cn}, Long Zhao$^{a,}$\footnote{zhaolong@jlu.edu.cn}, Zi-Xuan Zhao$^{a,}$\footnote{zzx23@mails.jlu.edu.cn
}

}
\vskip 2.5em
{\normalsize\it $^{a}$Center for Theoretical Physics and College of Physics, Jilin University,\\ Changchun 130012, People's Republic of China\\
$^{b}$Max Planck Institute for Gravitational Physics (Albert Einstein Institute),\\
Am M\"uhlenberg 1, 14476 Golm, Germany
}
\end{center}
\vskip 2.0em
\begin{abstract}
The fusion rules and operator product expansion (OPE) serve as crucial tools in the study of operator algebras within conformal field theory (CFT). Building upon the vision of using entanglement to explore the connections between fusion coefficients and OPE coefficients, we employ the replica method and Schmidt decomposition method to investigate the time evolution of entanglement entropy (EE) and pseudo entropy (PE) for linear combinations of operators in rational conformal field theory (RCFT). We obtain a formula that links fusion coefficients, quantum dimensions, and OPE coefficients. We also identify two definition schemes for linear combination operators. Under one scheme, the EE captures information solely for the heaviest operators, while the PE retains information for all operators, reflecting the phenomenon of pseudo entropy amplification. Irrespective of the scheme employed, the EE demonstrates a step-like evolution, illustrating the effectiveness of the quasiparticle propagation picture for the general superposition of locally excited states in RCFT. From the perspective of quasiparticle propagation, we observe spontaneous block-diagonalization of the reduced density matrix of a subsystem when quasiparticles enter the subsystem.
%We study the late-time behavior of R\'enyi entanglement entropy (REE) as well as pseudo R\'enyi entanglement entropy (PREE) in RCFT local operator quenches. The local operators are linear combinations of finite primary operators. We find that for generic linear combinations, the REE only captures the information of the heaviest primary, while the PREE preserves the information of all the primary operators, which reflects the effect of PE amplification. We also show that the maximum that can be reached by PREE but generally cannot be reached by REE perfectly fits the fusion rules in minimal models.
\end{abstract}
\baselineskip=0.7cm
\end{titlepage}
\tableofcontents
\section{Introduction}\label{section 1:intro}
In 2D conformal field theory (CFT), fusion rules determine which operators appear in the operator product expansion (OPE) of two primary fields, originating from the BPZ equations \cite{Belavin:1984vu, DiFrancesco:1997nk}. These rules are undoubtedly crucial since for 2D CFT all correlation functions can be reduced to the two-point function via the OPE. %On the other hand, the fusion algebra and fusion category derived from them play an important role in mathematics \cite{etingof2005fusion,drinfeld2010braided}, quantum computing \cite{wang2010topological,lahtinen2017short,rowell2018mathematics}, and condensed matter theories \cite{Feiguin:2006ydp,Barkeshli:2009tda,Kong:2014qka,Wen:2016ddy}.
One can summarize the fusion rules in the following commutative and associative fusion algebra\cite{Verlinde:1988sn,Fuchs:1993et}
\[
[\mo_i]\times[\mo_j]=&\sum_{k}N_{ij}^{k}[\mo_k],\label{fusionalgebra1}
\]
where  [$\mo_i$] denotes the conformal family of the primary $\mo_i$, and $N_{ij}^k$ called fusion numbers are non-negative integers. Note one can rephrase eqn.  \eqref{fusionalgebra1} in terms of the quantum dimension $d_i$ of the associated conformal family $[\mo_i]$ \cite{Verlinde:1988sn,fuchs1991quantum}
\[
d_id_j=\sum_k N_{ij}^kd_k.\label{fusion-quantum-dimension}
\]
 In rational conformal field theory (RCFT), it is well-known that the fusion numbers $N_{ij}^k$ can be computed using the Verlinde formula \cite{Verlinde:1988sn}. The counterpart in irrational conformal field theory (e.g., Liouville field theory)  was studied in \cite{Jego:2006ta}.

{Interestingly, recent studies on quantum entanglement in RCFT have seemingly offered a new perspective on the connection between fusion numbers and OPE coefficients \cite{Nozaki:2014hna, He:2014mwa, Caputa:2015tua,Chen:2015usa,Numasawa:2016kmo,Guo:2018lqq}. { The related topic is called  \textit{local operator quench} \cite{Nozaki:2014hna}, where one first generates a locally excited state through acting with a local operator $\mo$ on the vacuum state $|\Omega\rangle$ of the theory of interest, and then evolves it with the Hamiltonian of the theory,
\[
|\psi(t)\rangle:=\frac{1}{\sqrt{\mathcal{N}(\epsilon)}}e^{-iH(t-i\epsilon)}\mo(x)|\Omega\rangle,\label{teles}
\]
where $\epsilon$ is an infinitesimal positive parameter and $\mathcal{N}(\epsilon)$ is a normalization factor such that $\langle\psi(t)|\psi(t)\rangle=1$.} It has been shown that the local operator quench exhibits broad applicability in measuring scrambling and thermalization effects in  CFTs with large central charge \cite{Caputa:2014eta,Caputa:2014vaa,Asplund:2014coa,Asplund:2015eha,Caputa:2016tgt},  as well as in probing the bulk geometry \cite{Suzuki:2019xdq} and characterizing bulk dynamics \cite{Nozaki:2013wia,Caputa:2019avh,Kusuki:2019avm,Kawamoto:2022etl} in the context of AdS/CFT correspondence \cite{Maldacena:1997re,Gubser:1998bc,Witten:1998qj}. Investigating the local operator quench in RCFT is driven by the quest to understand the intricate connections between fusion rules, entanglement properties, and the broader implications for quantum information and computing. In particular, the fusion rules and their role in entanglement in CFT have connections to topological quantum computing \cite{wang2010topological,lahtinen2017short,rowell2018mathematics}. The entanglement dynamics obtained from such investigations contribute to theoretical advancements and potential practical applications.

{The process of local operator quench provides an excellent platform for the study of entanglement dynamics between subsystems. In RCFT local primary or descendant operator quench, i.e., the local operator $\mo$ in eqn. \eqref{teles} is a primary or descendant operator, it was found that the variations \footnote{  The variations of EE and REE, denoted as $\Delta S\big(\rho_A^{\mo}(t)\big)$ and  $\Delta S^{(n)}\big(\rho_A^{\mo}(t)\big)$ respectively in this paper, are defined as
\[
\Delta S\big(\rho^{\mo}_A(t)\big)\equiv \lim\limits_{n\to1}\Delta S^{(n)}\big(\rho^{\mo}_A(t)\big),\quad\Delta S^{(n)}\big(\rho^{\mo}_A(t)\big):=S^{(n)}\big(\rho_A^{\mo}(t)\big)-S^{(n)}\big(\rho_A^{\Omega}\big),
\]
where $\rho_A^{\mo}(t)$ is the reduced density matrix of subsystem $A$ when the total system is in the state $|\psi(t)\rangle$ \eqref{teles}, and $\rho_A^{\Omega}$ is the reduced density matrix for the vacuum state $|\Omega\rangle$.
} of entanglement entropy (EE, defined as $S(\rho_A):=-\tr[\rho_A\log\rho_A]$) and R\'enyi entanglement entropy (REE, defined as $S^{(n)}(\rho_A):=\frac{1}{1-n}\log\tr[\rho_A^n]$) saturate to a constant equal to the logarithm of quantum dimension of the local operator \cite{He:2014mwa, Caputa:2015tua, Chen:2015usa}. The result has been generalized to various cases \cite{Nozaki:2013wia
,Nozaki:2014uaa,Caputa:2014vaa,Caputa:2014eta,Guo:2015uwa,Caputa:2016tgt,Numasawa:2016kmo,He:2017lrg,Apolo:2018oqv,Guo:2018lqq,Kusuki:2019avm,Bhattacharyya:2019ifi,Kusuki:2019evw,Kawamoto:2022etl,Goto:2023wai,Kudler-Flam:2023ahk}, including the multiple local excitations \cite{Numasawa:2016kmo,Guo:2018lqq},
\[
|\Psi(t)\rangle=\frac{1}{\sqrt{\mathcal{N}(\epsilon)}}e^{-iH(t-i\epsilon)}\mo_N(x_N)...\mo_i(x_i)...\mo_1(x_1)|\Omega\rangle.\label{multiquench}
\]
In RCFT multiple local operator quenches,  the finiteness of quantum dimension guarantees the conservation of entanglement under the scattering of quasiparticles \cite{Numasawa:2016kmo}. When the subsystem $A$ is semi-infinite, the variations of the EE and REE at late times are solely the summation of the contributions from individual local operator quenches, i.e.,
\[
\lim_{t\to\infty}\Delta S^{(n)}(\rho^\Psi_A(t))=\sum_{i=1}^{N}\lim_{t\to\infty}\Delta S^{(n)}(\rho^{\mo_i}_A(t))=\sum_{i=1}^{N}\log d_i,\label{mutipleEE}
\]
where $d_i$ is the quantum dimension of $\mo_i$ \cite{Numasawa:2016kmo,Guo:2018lqq}.}

{This paper mainly focuses on the local quenches of linear combination operators in RCFT. Specifically, our interest centers on cases wherein local operator $\mo$ in eqn. \eqref{teles} assumes the form }
%In RCFT local quenches, it has been observed that the variations of entanglement entropy (EE) and R\'enyi entanglement entropy (REE) saturate to a constant value, which is equal to the logarithm of the quantum dimension of the local operator used in the quench. The conservation of entanglement implies that the total amount of entanglement in the system remains unchanged during the local quench. Since \cite{Chen:2015usa, He:2023eap} found that the variations of REE have been smeared, we are motivated to restore the conservation of entanglement in the local quench.
%This paper mainly focuses on the local quenches of linear combination operators in RCFT. The locally excited states of interest are of the form
\[
{\mo(x)=\sum_{k,\{\a\},\{\bar\a\}}C^{\{\alpha,\bar\alpha\}}_k\cdot L^{2h_k+|\a|+|\bar\a|}\cdot\mo^{\{\a,\bar\a\}}_k(x).}\label{linear-combination-operator}
\]
{In the above,  $C^{\{\a,\bar\a\}}_k$ are dimensionless combination coefficients, $L$ is a constant with a dimension of length (such as the length of subsystem), and $\mo^{\{\a,\bar\a\}}_k(x)$ are descendant operators; $\mo^{\{\a,\bar\a\}}_k(x)\equiv L_{-\a_1}...\bar L_{-\bar\a_1}...\mo_k(x)$,
$|\a|\equiv\sum_i\a_i$. Generically, in RCFTs, when the subsystem $A$ is semi-infinite, one may expect that the variation of the EE of operators taking the form of \eqref{linear-combination-operator} at late times should be a function of $C_{k}^{\{\a,\bar\a\}}$ and $d_k$, denoted as $\log (f(C_{k}^{\{\a,\bar\a\}}, d_k))$.}

The motivation for studying such kinds of local operator quenches will be clear from the perspective of OPE. The  OPE of two spin-zero primary fields $\mo_{i}$ takes the following form\footnote{We assume that $x_{1}$ and $x_{2}$ lie on the real axis.}
\[
\mo_i(x_1)\mo_j(x_2)=&\sum_{k}C_{kij}\sum_{\{\alpha\},\{\bar{\alpha}\}}\beta_{ij}^{k\{\a\}}\bar\beta^{k\{\bar\a\}}_{ij}|x_1-x_2|^{|\alpha|+|\bar\alpha|+2h_{k}-2h_{i}-2h_j}\mo^{\{\alpha,\bar\alpha\}}_k(x_2).\label{OPE}
\]
{ One can utilize the operators on both sides of the OPE \eqref{OPE} to construct locally excited states. Assume that the subsystem $A$ is semi-infinite. For multiple local excitations \eqref{multiquench} generated by the composite operator on the l.h.s. of the OPE \eqref{OPE}, we know from eqn. \eqref{mutipleEE} that the variation of the EE  (as well as REE) of the multiple locally excited state will eventually saturate to $\log d_i+\log d_j=\log d_id_j=\log \sum_{k}N_{ij}^k d_k$, where the second equality comes from the fusion rule of $\mo_i$ and $\mo_j$ (see eqn. \eqref{fusion-quantum-dimension}). On the other hand, given that the linear combination operator on the r.h.s. of the OPE \eqref{OPE} is solely a specific instance of eqn. \eqref{linear-combination-operator}, and based on the discussion following eqn. \eqref{linear-combination-operator}, we conclude that the variation of the EE of the operator on the r.h.s. of the OPE \eqref{OPE} saturate to $\log\big(f(C_{kij}\beta_{ij}^{k\{\a\}}\bar\beta_{ij}^{k\{\bar\a\}}, d_k)\big)$. Whether starting from the l.h.s. of the equation \eqref{OPE} or the right side, the results for the variation of the EE obtained should be the same. The function $f$ thus characterizes a connection between the OPE coefficients and fusion numbers in RCFT since, from the above discussions, we have $f(C_{kij}\beta_{ij}^{k\{\a\}}\bar\beta_{ij}^{k\{\bar\a\}}, d_k)=\sum_{k}{N}_{ij}^kd_k$. This is the primary motivation for studying the EE and REE associated with the linear combination operator \eqref{linear-combination-operator} in this paper.}

Even in Ising CFT, however, employing the standard replica method \cite{Callan:1994py, Calabrese:2004eu} to calculate the EE of a linear combination operator containing numerous descendant fields proves to be intricate (see \cite{Chen:2015usa} for a particular combination of descendant fields). Hence, here, we begin with a simplified scenario, excluding all descendant operators from \eqref{linear-combination-operator}\footnote{Note that two operators up to a global factor yield the same state. In this paper,  we partially fix this redundancy by imposing the normalization constraint $\sum_{k\in\S}|C_k|^2=1.$},
\[
|\psi\rangle=\frac{1}{\sqrt{\mathcal{N}(\epsilon)}}e^{-\epsilon H}\mo(x)|\Omega\rangle,\quad\mo(x)=\sum_{k\in\S}C_k\cdot L^{2h_k}\cdot\mo_k(x).\label{simplified-linear-combination-operator}
\]
where $\S$ is a primary operator set $\mathcal{S}:=\{\mo_1,\mo_2,...,\mo_m\}~(m\in\mathbb{Z}_+)$ from a RCFT.
Additionally, we assume that the operators $\mo_k$ in $\S$ are orthogonal to each other in terms of the two-point function, such that $\langle\mo^\dagger_k\mo_l\rangle\propto\delta_{kl}$. {Although such a blunt truncation overlooks the contributions of all descendant fields to entanglement, we anticipate that our analysis of the linear combination operator \eqref{simplified-linear-combination-operator} will provide valuable clues for the subsequent general discussions. {We will employ the replica method and the Schmidt decomposition method \cite{Guo:2018lqq} to compute the late-time EE of the linear combination operator \eqref{simplified-linear-combination-operator}. The latter method is a quantum mechanical approach that circumvents the complicated computations of all higher-point correlation functions. The consistency in results obtained from both methods instills confidence in utilizing the latter to compute the EE of operators on the r.h.s of the OPE \eqref{OPE}.

On the other hand,} we expect that our simple linear combination operators \eqref{simplified-linear-combination-operator} can reproduce the fusion relations \eqref{fusion-quantum-dimension} in RCFTs in a heuristic manner. Our strategy is as follows: starting from a known fusion equation \eqref{fusionalgebra1},  we select all primary operators appearing on the r.h.s. of it to construct the linear combination operator in \eqref{simplified-linear-combination-operator}.  {Subsequently, we compute the EE (REE) for the state \eqref{simplified-linear-combination-operator}.} It will be a function of the combination coefficients $\{C_k\}$ and operator quantum dimensions $\{d_k\}$. We look for the possibility in the parameter space of $\{C_k\}$ that operator \eqref{simplified-linear-combination-operator} produces the same EE (i.e., $\log\sum_{k}N^{k}_{ij} d_k$) as the linear combination operator on the r.h.s. of OPE \eqref{OPE} corresponding to the given fusion equation.
%Initially, we compute the EE and REE using the state \eqref{simplified-linear-combination-operator}. Subsequently, we determine the combinatorial coefficient $C_k$ by maximizing the variation of the EE (REE). To reproduce eqn.~\eqref{fusion-quantum-dimension} from the EE (REE), it is essential that the EE (REE) captures contributions from all local primary operators. To achieve this, we consider the late-time limit of the variation of the EE (REE). In this scenario, we easily derive the function $f$.
It is intriguing that, however, when attempting this, it can be proven that the EE and REE for the state \eqref{simplified-linear-combination-operator} only capture information about the heaviest operators (those with the highest conformal dimensions) in $\S$. For the most extreme situation, consider the { case in which the }set includes a unique heaviest primary operator $\mathcal{O}_M$, the expression for the late-time REE is
\[
\lim_{t\to\infty}\Delta S^{(n)}(\rho^\psi_{A}(t))=\log d_M,\label{only-logdm}
\]
where $d_M$ is the quantum dimension of $\mathcal{O}_M$, and its conformal dimension is denoted by $h_M = \max_{k\in\mathcal{S}}\{h_k\}$. In this situation, information regarding all operators apart from $O_M$ is entirely lost.
%One can also consider another extreme scenario, where all primary operators in $\S$ share the same conformal dimension. In such a case, the EE (REE) would contain the quantum dimensions of all operators \cite{Zhang:2019kwu}.  It can be proven that, up to a phase uncertainty, there exists a unique set of coefficients $\{C_k\}$ such that the late-time EE (REE) of the linear combination operator \eqref{simplified-linear-combination-operator} satisfies the fusion relation (please refer to section \ref{EEandREE} for more details). It is crucial to emphasize that for the general case where the conformal dimensions of operators in set $\S$ are not identical, we are unable to reconstruct the fusion relation from the EE (REE) of linear combination operators in \eqref{simplified-linear-combination-operator}.
%However, under general circumstances, we cannot reconstruct relation \eqref{fusion-quantum-dimension} from the late-time limit of the variation of the EE (REE).

To reproduce relation \eqref{fusion-quantum-dimension} using our simple linear combination operators \eqref{simplified-linear-combination-operator}, let us momentarily set aside the EE and replace it with {a generalisation version}, known as pseudo entropy (PE). PE is a two-state generalization of EE recently proposed via AdS/CFT and post-selection processes \cite{Nakata:2020luh}. It is defined as follows,
\[
S(\mathcal{T}^{\psi|\varphi}_A)=-\tr[\T_A^{\psi|\varphi}\log\T_A^{\psi|\varphi}].
\]
The operator $\T_A^{\psi|\varphi}$ involving two  states (i.e., the initial state $\psi$ and the final state $\varphi$ in the context of post-selection experiments \cite{aharonov2008two,RevModPhys.86.307}), called reduced transition matrix, is the partial trace of the transition matrix $\T^{\psi|\varphi}$ \cite{Nakata:2020luh,Guo:2022jzs},
\[
\T^{\psi|\varphi}=\frac{|\psi\rangle\langle\varphi|}{\langle\varphi|\psi\rangle}=\frac{\rho^\psi\rho^\varphi}{\tr[\rho^\psi\rho^\varphi]},\quad \T_A^{\psi|\varphi}=\tr_{\bar A}[T^{\psi|\varphi}].
\]
Like the calculation of EE, one usually computes the so-called pseudo-R\'enyi entanglement entropy (PREE) first. Subsequently, the PE is obtained by taking the limit of $n\to1$ to the $n$-th PREE,
\[
S^{(n)}(\mathcal{T}^{\psi|\varphi}_A)=\frac{1}{1-n}\log\tr[(\T_A^{\psi|\varphi})^n],\quad S(\mathcal{T}^{\psi|\varphi}_A)=\lim\limits_{n\to1}S^{(n)}(\mathcal{T}^{\psi|\varphi}_A).
\]
In previous studies \cite{Guo:2022sfl,He:2023eap
}, it was shown that, for subsystem $A$ being half-space, the $n$-th PREE under local primary or descendant operator quenches converges to the logarithm of the quantum dimension of the local operator as time $t$ approaches infinity.\footnote{See \cite{Wang:2018jva,Mollabashi:2020yie,Mollabashi:2021xsd,Goto:2021kln,Miyaji:2021lcq,Akal:2021dqt,Nishioka:2021cxe,Li:2022tsv,Ishiyama:2022odv,Doi:2022iyj,Doi:2023zaf,He:2023wko,Chen:2023gnh,Chen:2023eic,Guo:2023aio,Parzygnat:2023avh,He:2023ubi,Chu:2023zah,Jiang:2023loq,Jiang:2023ffu,Narayan:2023ebn,Kawamoto:2023nki,Omidi:2023env,Guo:2023tjv,Shinmyo:2023eci,Narayan:2023zen,Narayan:2022afv,Kanda:2023jyi} for more developments on PE and related topics.} In this work, to construct the transition matrix $\T^{\psi|\tilde\psi}$, we choose the state in eqn. \eqref{simplified-linear-combination-operator} as $|\psi\ra$ and select another state $|\tilde{\psi}\rangle$ as
\[
|\tilde\psi\rangle={\mathcal{N}(\epsilon)}^{-1/2}e^{-\epsilon H}\mo(\tilde{x})|\Omega\ra\,.\label{psi-tilde}
\]
As outlined in section \ref{section 2.1}, the advantage of the PE of the reduced transition matrix $\T^{\psi|\tpsi}_A$ compared with EE is its capability to retain information about all primary operators within $\mathcal{S}$. Specifically, the maximal value attainable by $S^{(n)}(\T_A^{\psi|\tpsi})$ but not achievable by $S^{(n)}(\rho_A^{\psi})$ in general is the logarithm of the sum of the quantum dimensions of all primary operators,
\[
\max_{\{C_k\}}\left\{\lim_{t\to\infty}\Delta S^{(n)}(\mathcal{T}_A^{\psi|\tilde\psi}(t))\right\}=\log\left(\sum_{k\in\S}d_k\right).\label{maximal-value-of-pseudo-entropy}
\]
Comparing \eqref{maximal-value-of-pseudo-entropy} with \eqref{only-logdm}, we observe that the late-time value of PE is larger than the EE, a phenomenon identified in \cite{Ishiyama:2022odv} as the \textit{pseudo entropy amplification}.

{The PE enables us to reproduce the fusion equation \eqref{fusion1} in RCFTs under the rough truncation from eqn. \eqref{linear-combination-operator} to eqn. \eqref{simplified-linear-combination-operator}. Consider, for instance, one of the fusion equations in the critical Ising model:
\[
\sigma\times\sigma=\mathbb{I}+\varepsilon,\label{is1}
\]
where $\sigma$ is the spin operator, $\varepsilon$ is the energy density operator and $\mathbb{I}$ the identity. The quantum dimensions of three primaries are $d_{\sigma}=\sqrt{2}$, $d_{\varepsilon}=d_{\mathbb{I}}=1$ respectively. We will analyze the late-time PE of the transition matrices $\mathcal{T}^{\psi|\tilde{\psi}}$ generated by both sides of eqn. \eqref{is1}:  the composite operator $\sigma(x+L)\sigma(x)$ and the linear combination operator $C_{\mathbb{I}}\cdot \mathbb{I}+C_{\varepsilon}\cdot L^{2h_{\varepsilon}}\cdot\varepsilon(x)$. For the composite operator $\sigma(x+L)\sigma(x)$, its late-time PE equals its late-time EE \cite{Guo:2022sfl}. Hence, according to eqn. \eqref{mutipleEE}, the value of the late-time PE we arrive at is $\log d_{\sigma}+\log d_{\sigma}=\log d^2_{\sigma}=\log2$.  On the other hand, considering the set $\S=\{\mathbb{I},\varepsilon\}$, according to eqn. \eqref{maximal-value-of-pseudo-entropy}, we find that the maximum value of the PE precisely equals $\log 2=\log (d_{\mathbb{I}}+d_{\varepsilon})$. Thus, we can always identify a set of appropriate combination coefficients $\{C_{\mathbb{I}}, C_{\varepsilon}\}$ to reproduce the fusion equation $d_{\sigma}\times d_{\sigma}=d_{\mathbb{I}}+d_{\varepsilon}$ through the PE.}

As previously mentioned, if we aim not only to reproduce fusion relations \eqref{fusion-quantum-dimension} using quantum entanglement but also to establish a connection between fusion coefficients and OPE coefficients, the contributions of all possible descendant operators in the OPE must be added. Although counting the contributions of an infinite number of operators seems implausible, Ref. \cite{Guo:2018lqq} presented an ingenious approach. {By imposing identical conformal transformations on both sides of the OPE \eqref{OPE}, the authors in \cite{Guo:2018lqq} deduced the contribution to EE from each conformal family that appears to the r.h.s. of the OPE \eqref{OPE}, i.e., the EE of the operator
\[
\sum_{\{\alpha\},\{\bar{\alpha}\}}\beta_{ij}^{k\{\a\}}\bar\beta^{k\{\bar\a\}}_{ij}|x_1-x_2|^{|\alpha|+|\bar\alpha|+2h_{k}-2h_{i}-2h_j}\mo^{\{\alpha,\bar\alpha\}}_k(x_2).\label{OPEcf}
\]
For RCFTs, this contribution precisely equates to the logarithm of the corresponding quantum dimension (i.e., $\log d_k$). Based on this nontrivial result, we can utilize the Schmidt decomposition method to derive the late-time EE of the linear combination operator on the r.h.s. of the OPE \eqref{OPE}. The result is a function of the OPE coefficients and quantum dimensions (see eqn. \eqref{e78}). As we mentioned in the discussions after eqn. \eqref{OPE}, this function characterizes a relationship between the OPE coefficients and the fusion coefficients because  we have $f(C_{kij}\beta_{ij}^{k\{\a\}}\bar\beta_{ij}^{k\{\bar\a\}}, d_k)=\sum_{k}{N}_{ij}^kd_k$. By means of the technique we used in exploring the maximum PE of the simplified linear combination operator \eqref{simplified-linear-combination-operator} (further details provided in section \ref{section:fusion and OPE}), we finally establish a link (see eqn. \eqref{formula:FusionAndOPE}) between fusion coefficients, quantum dimensions, and OPE coefficients using the function \eqref{e78}.}
%The result \eqref{maximal-value-of-pseudo-entropy} includes the quantum dimensions of all operators, so in principle, it is possible to reconstruct fusion relation \eqref{fusion-quantum-dimension}. For the minimal models (RCFT?), this method is very effective, and we can easily determine that their fusion numbers are either 0 or 1. This is consistent with the results obtained by carefully calculating the entanglement entropy through constructing the reduced density matrix with state \eqref{linear-combination-operator}, demonstrating the advantage of pseudo-entropy over entanglement entropy.

Although we reproduce the fusion relations by computing the PE of the linear combination operator \eqref{simplified-linear-combination-operator} and establish the correlation between fusion coefficients and OPE coefficients through the linear combination operator like \eqref{linear-combination-operator}, these operators might not be suitable when considering the EE of a finite number of linear combinations of operators. As mentioned earlier, the information about light operators doesn't manifest in the EE. %The reason lies in the fact that the superposition states obtained through operators like \eqref{simplified-linear-combination-operator} are not natural from a certain point of view. In such cases, the weights of states generated by light operators in $\mo(x)$ tend toward zero as the regulator $\e$ approaches zero (see section \ref{section2.3} for more details).
To find a way by which the information about light operators in $\S$ can be expressed within the EE is the second objective of this paper. It turns out that by considering the following refined linear combination operator
\[
\mo(x;\epsilon)=\sum_{k\in\S} C_k\frac{\mo_k(x)}{\sqrt{\la\mo_k^\dagger(x,\epsilon)\mo_k(x,-\epsilon)\ra}}=\sum_{k\in\S} C_k\cdot|2\epsilon|^{2h_k}\cdot\mo_k(x)\label{normalized-linear-combination-operator}
\]
instead of the linear combination operator in eqn. \eqref{simplified-linear-combination-operator}, all information of operators in $\S$ can be encoded into the late-time EE (see section \ref{section2.3} for more details). We can delve into further details of entanglement dynamics about operator $\eqref{normalized-linear-combination-operator}$, particularly regarding the applicability of the quasiparticles picture. It's well-known that each primary operator $\mo_k$ in \eqref{normalized-linear-combination-operator}  can generate a locally excited state with a classical interpretation of quasiparticles propagation \cite{Nozaki:2014hna,He:2014mwa}, whether the state generated by operator \eqref{normalized-linear-combination-operator} (i.e., the superposition of locally primary excited states) still possesses the quasiparticles picture is not immediately apparent. A common counterexample is that in quantum mechanics the superposition of coherent (quasiclassical) states usually no longer has a classical correspondence. Using the replica method, we find that the answer is positive. The evolution of EE (and REE) corresponding to operator $\eqref{normalized-linear-combination-operator}$ shows a step-like behavior from zero to non-zero, which suggests that the quasiparticle propagation picture indeed holds for the general superposition of locally excited states in RCFT. Another fascinating detail in entanglement dynamics is our observation that, under the quasiparticle propagation picture, upon the entry of quasiparticles into a subsystem, the reduced density matrix of the subsystem spontaneously undergoes instantaneous block-diagonalization. Each distinct diagonal block carries information about different conformal families.

%We investigate the time evolution of its EE and observe that its characteristics align with the quasiparticles propagation scenario. This finding is significant for states generated by the action of linear combinations of operators, i.e., superposition states. It suggests that the density matrix of such states is block-diagonalized according to the Verma module, as demonstrated in the subsequent subsection.

The rest of this paper is organized as follows.
In section \ref{section 2}, we derive the late-time REE and PREE of the operator \eqref{simplified-linear-combination-operator} through two methods, the Schmidt decomposition approach and the replica trick. We analyze the dependence of REE and PREE on the combination coefficients, showcasing the emergence of pseudo entropy amplification by demonstrating the maximum values of REE and PREE. Utilizing the Schmidt decomposition method, we establish a connection between fusion and OPE coefficients by examining the REE of operators on both sides of the OPE. In section \ref{section 3 discussion}, we illustrate how operator \eqref{normalized-linear-combination-operator} facilitates the restoration of information related to light operators in the EE (REE). Additionally, we present further details of the entanglement dynamics associated with operator \eqref{normalized-linear-combination-operator}.  The summary and prospect are given in section \ref{sec4}.
\vskip 0.4em

\section{Pseudo entropy amplification and fusion rules}\label{section 2}

The earliest exploration of the EE (REE) of a superposition of locally primary excited states appeared in \cite{Guo:2018lqq}, followed by an extension of the PE (PREE) by the authors of \cite{Guo:2022sfl}. According to the Schmidt decomposition method \cite{Guo:2018lqq}, the late-time limit of the $n$-th PREE of the reduced transition matrix generated by operator \eqref{simplified-linear-combination-operator} is given by \cite{Guo:2022sfl}
\[
\lim_{t\to\infty}\Delta S^{(n)}(\T^{\psi|\tilde\psi}_A(t))
=&\frac{1}{1-n}\log\left[\sum_{k\in\S}\left(\frac{d_k^{\frac{1}{n}-1}|C_k|^2L^{4h_k}\langle\mathcal{O}^\dagger_k(\tilde x,\epsilon)\mathcal{O}_k(x,-\epsilon)\rangle}{\sum_{j\in\S}|C_{j}|^2L^{4h_{j}}\langle\mathcal{O}^\dagger_{j}(\tilde x,\epsilon)\mathcal{O}_{j}(x,-\epsilon)\rangle}\right)^n\right].\label{formula-latetimePREE-sec2.1}
\]
Setting $\tilde x=x$ allows one to obtain the late-time EE associated with operator \eqref{simplified-linear-combination-operator}. Despite the good agreement between formula \eqref{formula-latetimePREE-sec2.1} and numerical computations, an analytical derivation based on the standard method of the CFT replica trick remains absent to date. In this section, we bridge this gap by providing replica calculations for the cases of $n=2$ and $n=3$, and the computation for general $n$ can be obtained straightforwardly. Subsequently, we aim to extract intriguing information from the late-time PREE formula. We show the pseudo entropy amplification phenomenon by exploring
the maximum values of PREE and REE of locally excited states \eqref{simplified-linear-combination-operator}. These research findings point towards a potential relationship between fusion and OPE coefficients as demonstrated in subsection \ref{section:fusion and OPE}.

%{The late-time PREE was first derived by employing the Schmidt decomposition method \cite{Guo:2018lqq} rather than the standard CFT replica trick. In this section, we bridge this gap by providing replica calculations for the cases of $n=2$ and $n=3$, and the computation for general $n$ can be obtained straightforwardly. Then we show the pseudo entropy amplification phenomenon by exploring
%the maximum values of PREE and REE of locally excited states \eqref{simplified-linear-combination-operator}. These research findings point towards a potential relationship between fusion coefficients and OPE coefficients as demonstrated in subsection \ref{section:fusion and OPE}.}
%More importantly, we will demonstrate that more interesting information can be extracted from eqn. \eqref{nthPEE-linear-combination-operator}. We also present a corresponding formula for the R\'enyi entanglement entropy (REE), beyond simply removing the wavy line from the above PREE formula. We summarize our main results as follows:
\subsection{The late-time formula of $n$-th PREE/REE}\label{section 2.1}
{ In replica method, calculating the $n$-th REE/PREE for an operator $\mo$, which consists of a linear combination of $m$ primary operators, requires working with $m^{2n}$ $2n$-point functions on the plane. {In this subsection, we first review the derivation of eqn. \eqref{formula-latetimePREE-sec2.1} by Schmidt composition method in Ref. \cite{Guo:2022sfl}, and then  we will demonstrate the replica calculations of the second and third REE and PREE to derive the same results \eqref{formula-latetimePREE-sec2.1}.} %{\color{}We first  review the method of Schmidt composition, as introduced in \cite{Guo:2018lqq}, }}
\subsubsection{Schmidt decomposition method}
 As mentioned earlier, operators in $\S$ are mutually orthogonal in the sense of the two-point function $\la\mo^\dagger_k\mo_l\ra$. Let us first define a series of normalized excited states with the help of $\mathcal O_k\in\S$,
\begin{align}
|\mathcal{O}_k(x,t)\rangle:=&\frac{1}{\sqrt{\langle\mathcal{O}_k^{\dagger}(x,\epsilon)\mathcal{O}_k(x,-\epsilon)\rangle}}e^{-iH t}\mathcal{O}_k(x,-\epsilon)|\Omega\rangle,\quad \langle\mo_{k}(x,t)|\mo_{l}(x,t)\rangle=\delta_{kl}.\label{Opstate}
\end{align}
Generally speaking, $|\mathcal{O}_k(x,t)\rangle$ is an entangled state living in the sub-Verma module $V_k\bigotimes \bar V_{\bar{k}}$.  It can be written in the following form by a time-dependent Schmidt decomposition
\[
|\mathcal{O}_k(x,t)\rangle=\sum_ua^k_u(x,t)|k_u(x,t)\rangle\otimes|\bar k_u(x,t)\rangle,
\]
where $\{|k_u(x,t)\rangle\}$ and $\{|\bar{k}_u(x,t)\rangle\}$ parameterized by $(x,t)$  are two orthonormal bases of $V_k$ and $\bar V_{\bar{k}}$  repectively, and $a^k_u(x,t)\geq0$ $(\sum_u (a^k_u(x,t))^2=1)$ are Schmidt coefficients. The REE between $V_k$ and $\bar V_{\bar k}$ for the state $|\mo_k(x,t)\rangle$ is directly given by
\[
S^{(n)}\big(\rho^{\mo_k}_{V_k}(t)\big)=&\frac{1}{1-n}\log\big\{\Tr_{V_k}\big[\left(\Tr_{\bar V_{\bar{k}}}|\mathcal{O}_k(x,t)\rangle\langle\mathcal{O}_k(x,t)|\right)^n\big]\big\}\nn\\
=&\frac{1}{1-n}\log \sum_u\big(a^k_u(x,t)\big)^{2n}\label{compare1}.
\]
The next crucial point lies in the identification\footnote{This identification can be understood through the following physical intuition: In the quasiparticles propagation picture, in the large-$t$ limit, subsystem $A=[0,\infty)$ collects all the right-moving quasiparticles, which carry information about the holomorphic part of $\mo_k$, while the complement $\bar A$ collects all the left-moving quasiparticles, which carry information about the anti-holomorphic part of $\mo_k$ \cite{Nozaki:2014hna} \cite{He:2024tba} } of $\Delta S^{(n)}\big(\rho^{\mo_k}_A(t)\big)$ and $S^{(n)}\big(\rho^{\mo_k}_{V_k}(t)\big)$ in the large-$t$ limit, where
\[
\rho_A^{\mo_k}(t)\equiv\tr_{\bar A}\big[|\mo_k(x,t)\rangle\langle \mo_k(x,t)|\big]\label{rhoak}
\]
representing the reduced density matrix of subsystem $A=[0,\infty)$ when the total system is quenched by $\mo_k(x)$ at the initial time.  Consequently,
we arrive at the limit as $t\to\infty$ \cite{He:2014mwa}
\[
\lim\limits_{t\to\infty}S^{(n)}\big(\rho^{\mo_k}_{V_k}(t)\big)=\log d_k,
\]
where $d_k$ is the quantum dimension of $\mo_k$.
Now we move to investigate the PREE of $e^{-iHt}|\psi\rangle$ \eqref{simplified-linear-combination-operator} and $e^{-iHt}|\tilde\psi\rangle$ \eqref{psi-tilde}. These two states can be rewritten as two superposition states of $|\mathcal{O}_k(x,t)\rangle$
\[
e^{-iHt}|\psi\rangle=\sum_{k\in\S}\sqrt{\lambda_k}|\mathcal{O}_k(x,t)\rangle,\quad e^{-iHt}|\tilde\psi\rangle=\sum_{k\in\S}\sqrt{\tilde\lambda_k}|\mathcal{O}_k(\tilde x,t)\rangle,\label{psi(t)-linear-combination}
\]
where
\begin{align}
\lambda_k=\tilde\lambda_k=\frac{(C_k)^{2}{L^{4h_{k}}\langle\mathcal{O}_k^\dagger(x,\epsilon)\mathcal{O}_k(x,-\epsilon)\rangle}}{\sum_{j\in\S}|C_{j}|^2L^{4h_{j}}\langle\mathcal{O}_{j}^\dagger(x,\epsilon)\mathcal{O}_{j}(x,-\epsilon)\rangle},\quad\sum_{k\in\S}|\lambda_k|=1.\label{lambdap}
\end{align}
Consequently, the transition matrix becomes
\begin{align}
\mathcal{T}^{\psi|\tilde\psi}(t)=&\frac{1}{\sum_{k\in\S}\sqrt{\lambda_k}\sqrt{\tilde\lambda^*_k}\langle\mathcal{O}_k(\tilde x,t)|\mathcal{O}_k(x,t)\rangle}\nn\\
&\times\sum_{j,l\in\S}\sqrt{\lambda_j}\sqrt{\tilde\lambda^*_{l}}\sum_{u,v}a^j_u(x,t)a^{l}_v(\tilde x,t)|j_u(x,t)\rangle|\bar j_u(x,t)\rangle\langle l_v(\tilde x,t)|\langle\bar l_v(\tilde x,t)|.
\end{align}
The reduced transition matrix is obtained by tracing out the anti-holomorphic part ($V\equiv\oplus_k V_k$),
\begin{align}
\mathcal{T}_V^{\psi|\tilde\psi}(t)=&\Tr_{\oplus\bar k \bar V_{\bar k}}\mathcal{T}^{\psi|\tilde\psi}(t)\nonumber\\
=&\sum_{k\in\S}\sum_{u,v,w}\frac{\sqrt{\lambda_k}\sqrt{\tilde\lambda^*_{k}}a^k_u(x,t)a^{k}_v(\tilde x,t)\langle \bar k_v(\tilde x,t)|\bar k_{u}(x,t)\rangle\langle k_v(\tilde x,t)|k_w(x,t)\rangle}{\sum_{j\in\S}\sqrt{\lambda_{j}}\sqrt{\tilde\lambda^*_{j}}\langle\mathcal{O}_{j}(\tilde x,t)|\mathcal{O}_{j}(x,t)\rangle}\cdot|k_u(x,t)\rangle\langle k_w(x,t)|\label{uneasy},
\end{align}
which in general is nondiagonal under the basis $\{|k_u(x,t)\rangle\}$. We can compute the trace of $(\mathcal{T}_{V}^{\psi|\tilde\psi}(t))^n$,
\begin{align}
&\Tr\left[(\mathcal{T}_{V}^{\psi|\tilde\psi}(t))^n\right]\nn\\
=&\sum_{k\in\S}\frac{\Big(\sqrt{\lambda_k}\sqrt{\tilde\lambda^*_{k}}\Big)^n}{\Big(\sum_{j\in\S}\sqrt{\lambda_{j}}\sqrt{\tilde\lambda^*_{j}}\langle\mathcal{O}_{j}(\tilde x,t)|\mathcal{O}_{j}(x,t)\rangle\Big)^n}\nonumber\\
&\times\sum_{\{u\},\{v\} }a^k_{u_1}(x,t)a^{k}_{v_1}(\tilde x,t)\langle \bar k_{v_1}(\tilde x,t)|\bar k_{u_1}(x,t)\rangle\langle k_{v_1}(\tilde x,t)|k_{u_2}(x,t)\rangle...a^k_{u_n}(x,t)a^{k}_{v_n}(\tilde x,t)\nn\\
&\times\langle \bar k_{v_n}(\tilde x,t)|\bar k_{u_n}(x,t)\rangle\langle k_{v_n}(\tilde x,t)|k_{u_1}(x,t)\rangle\label{tobereduce}.
\end{align}
To further reduce $\eqref{tobereduce}$, let us turn to consider the $n$-th PREE of $|\mo_k(x,t)\ra$ and $|\mo_k(\tilde x,t)\ra$.
\[
S^{(n)}\big(\T_V^{\mo_k(x)|\mo_k(\tilde x)}(t)\big)=&\frac{1}{1-n}\log\Tr_{\oplus_kV_k}\Big[\big(\mathcal{T}_{V}^{\mo_k(x)|\mo_k(\tilde x)}(t)\big)^n\Big]\nonumber\\
=&\frac{1}{1-n}\log\Big[\langle\mathcal{O}_k(\tilde x,t)|\mathcal{O}_k(x,t)\rangle^{-n}\nonumber\\
&\times\sum_{\{u\},\{v\} }a^k_{u_1}(x,t)a^k_{v_1}(\tilde x,t)\langle \bar k_{v_1}(\tilde x,t)|\bar k_{u_1}(x,t)\rangle\langle k_{v_1}(\tilde x,t)|k_{u_2}(x,t)\rangle...a^k_{u_n}(x,t)a^k_{v_n}(\tilde x,t)\nn\\
&\times\langle \bar k_{v_n}(\tilde x,t)|\bar k_{u_n}(x,t)\rangle\langle k_{v_n}(\tilde x,t)|k_{u_1}(x,t)\rangle\Big].\label{compare2}
\]
Under the same spirit, we expect that
\[
\lim_{t\to\infty}\Delta S^{(n)}\big(\T^{\mo_k(x)|\mo_k(\tilde x)}_A(t)\big)=\lim_{t\to\infty}S^{(n)}\big(\T_V^{\mo_k(x)|\mo_k(\tilde x)}(t)\big).
\]
We have known that the late-time limit of $\Delta S^{(n)}\big(\mathcal{T}_A^{\mo_k(x)|\mo_k(\tilde x)}(t)\big)$ is equal to the late-time limit of $\Delta S^{(n)}\big(\rho^{\mo_k}_A(t)\big)$ \cite{Guo:2022sfl}, and the latter we already know is equal to \eqref{compare1}.

Comparing eqn. $\eqref{compare1}$ with eqn. $\eqref{compare2}$, we obtain the equality
\begin{align}
\lim_{t\to\infty}&\sum_{\{u\},\{v\} }a^k_{u_1}(x,t)a^k_{v_1}(\tilde x,t)\langle \bar k_{v_1}(\tilde x,t)|\bar k_{u_1}(x,t)\rangle\langle k_{v_1}(\tilde x,t)|k_{u_2}(x,t)\rangle...a^k_{u_n}(x,t)a^k_{v_n}(\tilde x,t)\nn\\
&\times\langle \bar k_{v_n}(\tilde x,t)|\bar k_{u_n}(x,t)\rangle\langle k_{v_n}(\tilde x,t)|k_{u_1}(x,t)\rangle\nonumber\\
=&\langle\mathcal{O}_k(\tilde x,t=0)|\mathcal{O}_k(x,t=0) \rangle^n d_k^{1-n}\label{equality}
\end{align}
Substituting eqn. $\eqref{equality}$ into eqn. $\eqref{tobereduce}$ and taking $t\to\infty$,  we finally obtain the late-time PREE formula
\[
\lim_{t\to\infty}\Delta S^{(n)}(\T^{\psi|\tilde\psi}_A(t))=&\lim_{t\to\infty}\frac{1}{1-n}\log\Tr\left[(\mathcal{T}_{V}^{\psi|\tilde\psi}(t))^n\right]\nn\\
=&\frac{1}{1-n}\log\left[\sum_{k\in\S}\left(\frac{d_k^{\frac{1}{n}-1}|C_k|^2L^{4h_k}\langle\mathcal{O}^\dagger_k(\tilde x,\epsilon)\mathcal{O}_k(x,-\epsilon)\rangle}{\sum_{j\in\S}|C_{j}|^2L^{4h_{j}}\langle\mathcal{O}^\dagger_{j}(\tilde x,\epsilon)\mathcal{O}_{j}(x,-\epsilon)\rangle}\right)^n\right].\label{resultsfinalapp}
\]
The accuracy of the above formula is verified by comparing it with the results obtained through numerical replica calculations in \cite{Guo:2022sfl}. The next section will employ the replica method and conformal mapping in 2D CFT to reproduce eqn. \eqref{resultsfinalapp} for $n=2$ and $n=3$.
\subsubsection{Replica method}\label{section: replica method}
Let us begin with the Euclidean path integral formulation of the replica calculation of the $n$-th PREE. Suppose that the theory of interest (with a Lagrangian $\L(\phi,\pd\phi)$) resides on a Euclidean plane $\Sigma_1$ with the metric $ds^2=dwd\bar w$ ($(w,\bar w)=(x+i\tau, x-i\tau)$). The Euclidean transition matrix of interest is generated by the linear combination operator $\mo(w,\bar w)\equiv e^{\tau H}\mo(x) e^{-\e H}$ \eqref{simplified-linear-combination-operator}
\[
\T^{\psi|\tilde\psi}_{E}=\frac{\mo(w_1,\bar w_1)|\Omega\ra\la\Omega|\mo^\dagger(w_2,\bar w_2)}{\la\Omega|\mo^\dagger(w_2,\bar w_2)\mo(w_1,\bar w_1)|\Omega\ra},
\]
where $w_1\equiv x-i\tau_1$, $w_2\equiv \tilde x+i\tau_2$ ($\tau_{1,2}>0$). The reduced transition matrix of
subsystem $A$ can be expressed by path integral with operators inserted at $(w_1,\bar w_1)$ and $(w_2,\bar w_2)$ on the $w$-plane with a cut on $A$
\begin{align}
\langle \phi_{A_-}|\mathcal{T}^{\psi|\tilde\psi}_{E,A}|\phi_{A+}\rangle=&\frac{\int^{\phi(x\in A,\tau=0_+)=\phi_{A_+}(x)}_{\phi(x\in A,\tau=0_-)=\phi_{A_-}(x)}[d\phi]\mo^{\dagger}(w_2,\bar w_2)\mo(w_1,\bar w_1)\exp\left\{-\int_{\mathbb{R}^2}\mathcal{L}(\phi,\partial\phi)\right\}}{\int[d\phi]\mo^{\dagger}(w_2,\bar w_2)\mo(w_1,\bar w_1)\exp\left\{-\int_{\mathbb{R}^2}\mathcal{L}(\phi,\partial\phi)\right\}}\nn\\
=&\left(\imineq{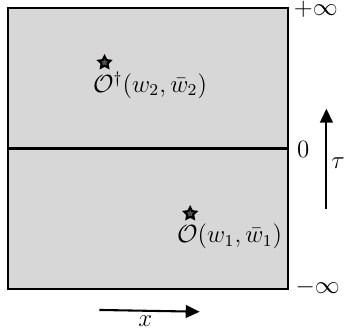}{16}\right)^{-1}\times~~~\imineq{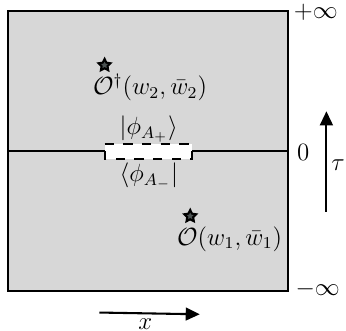}{16}.\label{reduced transition matrix A: graph}
\end{align}
Based on \eqref{reduced transition matrix A: graph}, $\tr[(\mathcal{T}^{\psi|\tilde\psi}_{E,A})^n]$  amounts to a $2n$-point correlation function on a $n$-sheet Riemann surface $\Sigma_n$,
\begin{align}
\tr[(\mathcal{T}_{E,A}^{\psi|\tilde\psi})^n]=&\left(\imineq{TA2.pdf}{16}\right)^{-n}\times~~~\imineq{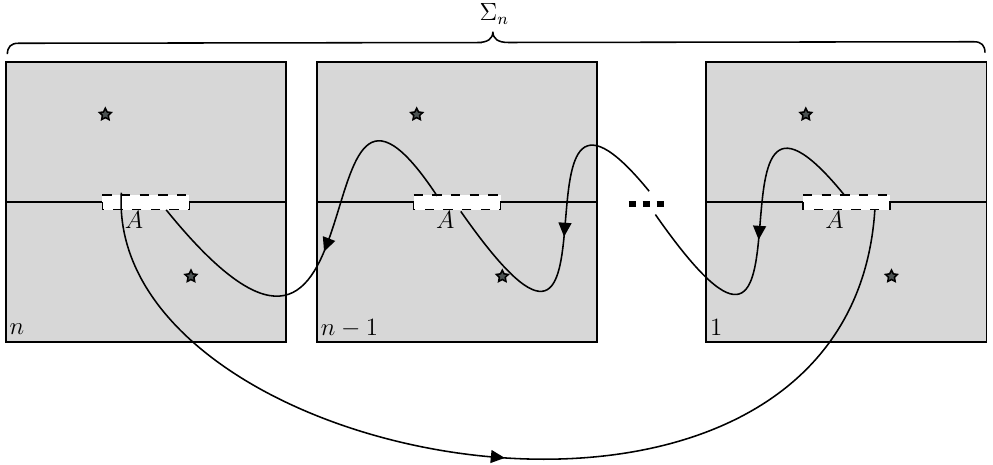}{22}\nn\\
=&\frac{Z_n}{Z_1^n}\cdot\frac{\langle \mo(w_1,\bar w_1) \mo^\dagger(w_2,\bar w_2)...\mo(w_{2n-1},\bar w_{2n-1})\mo^\dagger(w_{2n},\bar w_{2n})\rangle_{\Sigma_n}}{\langle \mo^\dagger(w_2,\bar w_2)\mo(w_1,\bar w_1)\rangle_{\Sigma_1}^n}\label{reduced transition matrix An: graph},
\end{align}
where $Z_1$ and $Z_n$ are vacuum partition functions on $\Sigma_1$ and $\Sigma_n$ respectively, and $\mo(w_{2k-1},\bar w_{2k-1})$ and $\mo^\dagger(w_{2k},\bar w_{2k})$ are operators inserted at the $k$-th sheet. The variation of the $n$-th PREE is defined  as subtracting the contribution of the $n$-th REE of the vacuum from the $n$-th PREE,
\[
\Delta S^{(n)}(\mathcal{T}^{\psi|\tilde\psi}_{E,A})\equiv& S^{(n)}(\mathcal{T}^{\psi|\tilde\psi}_{E,A})-S^{(n)}(\rho_{A,\text{vacuum state}})\nn\\
=&\frac{1}{1-n}\log \frac{\langle \mo(w_1,\bar w_1) \mo^\dagger(w_2,\bar w_2)...\mo(w_{2n-1},\bar w_{2n-1})\mo^\dagger(w_{2n},\bar w_{2n})\rangle_{\Sigma_n}}{\langle\mo(w_1,\bar w_1) \mo^\dagger(w_2,\bar w_2)\rangle_{\Sigma_1}^n}.
\]
The real-time evolution of the $n$-th PREE is achieved by performing a Wick rotation of the Euclidean time: $\tau_1\to\e+it$, $\tau_2\to\e-it$. For the case of $n=2$, $\Delta S^{(n)}(\mathcal{T}^{\psi|\tilde\psi}_{E,A})$ is linked to a series of four-point functions on $\Sigma_2$
\[
e^{-\Delta S^{(2)}(\mathcal{T}^{\psi|\tilde\psi}_{E,A})}=\frac{\sum_{ijkl\in\S}C_iC_j^*C_kC_l^*L^{2(h_i+h_j+h_k+h_l)}\la\mo_i(w_1,\bar w_1)\mo^\dagger_j(w_2,\bar w_2)\mo_k(w_3,\bar w_3)\mo^\dagger_l(w_4,\bar w_4)\ra_{\Sigma_2}}{\left(\sum_{j\in\S}|C_j|^2L^{4h_j}\la\mo_j(w_1,\bar w_1)\mo^\dagger_j(w_2,\bar w_2)\ra_{\Sigma_1}\right)^2}.\label{sec2-second-PREE}
\]
These four-point functions can be evaluated by applying the conformal mapping $z=i\sqrt{-w}$ \cite{Nozaki:2014hna,He:2014mwa} from   $\Sigma_2$ to $\Sigma_1$. The four operators after the mapping are located at
\[
&z_1=-z_3=i\sqrt{-x-t+i\e},\quad \bar z_1=-\bar z_3=-i\sqrt{-x+t-i\e},\nn\\
&z_2=-z_4=i\sqrt{-\tilde x-t-i\e},~~~\bar z_2=-\bar z_4=-i\sqrt{-\tilde x+t+i\e}.\label{z1234}
\]
Let's take $i=1$, $j=2$, $k=3$, and $l=4$ for the sake of writing convenience, the four-point function can be expressed as
\[
&\la\mo_1(w_1,\bar w_1)\mo^\dagger_2(w_2,\bar w_2)\mo_3(w_3,\bar w_3)\mo^\dagger_4(w_4,\bar w_4)\ra_{\Sigma_2}\nn\\
=&\prod_{i=1}^4|2z_i|^{-2h_i} \la\mo_1(z_1,\bar z_1)\mo^\dagger_2(z_2,\bar z_2)\mo_3(z_3,\bar z_3)\mo^\dagger_4(z_4,\bar z_4)\ra_{\Sigma_1}\nn\\
=&\left(\bigg|\frac{2z_1z_{12}z_{14}}{z_{24}}\bigg|^{-2h_1}\prod_{i=2}^4\bigg|\frac{2z_iz^2_{1i}z_{24}}{z_{12}z_{14}}\bigg|^{-2h_i}\right)G^{21}_{34}(\eta,\bar\eta),\label{O1234}
\]
where $G_{34}^{21}(\eta,\bar\eta)\equiv\lim_{z\to\infty}z^{2h_1}\bar{z}^{2h_1} \la\mo_1(z,\bar z)\mo^{\dagger}_2(1,1)\mo_3(\eta,\bar\eta)\mo^\dagger_4(0,0)\ra_{\Sigma_1}$, and $(\eta,\bar\eta)\equiv (\frac{z_{12}z_{34}}{z_{13}z_{24}},\frac{\bar z_{12}\bar z_{34}}{\bar z_{13}\bar z_{24}})$. Let us begin by coping with the $G$-function in terms of the conformal blocks. In general CFTs, $G^{21}_{34}(\eta,\bar\eta)$ can be written as follows \cite{Belavin:1984vu}
\[
G_{34}^{21}(\eta,\bar\eta)=\sum_{p}C_{34}^{p}C_{12}^pF^{21}_{34}(p|\eta)\bar{F}^{21}_{34}(p|\bar\eta). \label{conformal-block-expansion}
\]
Here, $C^p_{34}$ is the coefficient of the three-point function $\la\mo_p\mo_3\mo^\dagger_4\ra$, and the index $p$ corresponds to each $\mo_p$ of all Virasoro primary fields. It is important to note that the conformal block has a universal behavior around $\eta=0$,
\[
F_{34}^{21}(p|\eta)=\eta^{h_p-h_3-h_4}(1+O(\eta)).\label{conformal-block-at-0}
\]
As  shown in \cite{Guo:2022sfl}, the cross ratios $(\eta,\bar\eta)\simeq(1+\frac{(\tilde x-x+2i\epsilon)^2}{16t^2},-\frac{(\tilde x-x-2i\epsilon)^2}{16t^2})\sim(1,0)$ in the late-time limit. Then the fusion
transformation in RCFT \cite{Moore:1988ss,Moore:1988uz}
\begin{equation}
F^{21}_{34}(p|1-\eta)=\sum_{q}F_{pq}\left[\begin{smallmatrix}2,1\\3,4\end{smallmatrix}\right] F^{41}_{32}(q|\eta)
\end{equation}
can be leveraged to fix the leading behavior of the conformal block in $\eta\to1$ limit and subsequently fix the leading order of the late-time behavior of $G^{21}_{34}(\eta,\bar \eta)$
\[
\lim\limits_{t\to\infty}G_{34}^{21}(\eta,\bar\eta)\simeq&\sum_{p,q}C_{12}^{p} C_{34}^{p}F_{pq}\left[\begin{smallmatrix}2,1\\3,4\end{smallmatrix}\right](1-\eta)^{h_q-h_2-h_3}\bar\eta^{h_{p}-h_3-h_4}\nn\\
\sim&\sum_{p,q}C_{12}^{p} C_{34}^{p}F_{pq}\left[\begin{smallmatrix}2,1\\3,4\end{smallmatrix}\right] t^{2(h_2+2h_3+h_4-h_p-h_q)}.\label{late-time-G}
\]
On the other hand, we can determine the leading behavior of the prefactor for $G_{34}^{21}$ at late times by simply applying the expressions of $(z_i,\bar z_i)$ from eqn. \eqref{z1234}. It is found that
\[
\lim\limits_{t\to\infty}\left(\bigg|\frac{2z_1z_{12}z_{14}}{z_{24}}\bigg|^{-2h_1}\prod_{i=2}^4\bigg|\frac{2z_iz^2_{1i}z_{24}}{z_{12}z_{14}}\bigg|^{-2h_i}\right)\sim t^{-2(h_2+2h_3+h_4)}.\label{late-time-prefactor}
\]
Combining eqn. \eqref{late-time-G} with eqn. \eqref{late-time-prefactor} yields a remarkable outcome: the four-point function $\la\mo_1\mo^\dagger_2\mo_3\mo^\dagger_4\ra$ does not vanish at late times if, and only if, a vacuum sector is present in the fusion of any two operators
\[
\mo_1\times\mo_2=\mathbb{I}+...,&\quad \mo_3\times\mo_4=\mathbb{I}+...,\nn\\
\mo_1\times\mo_4=\mathbb{I}+...,&\quad \mo_2\times\mo_3=\mathbb{I}+... .\label{four-fusion}
\]
The only scenario that aligns with eqn. \eqref{four-fusion} is when $\mo_1=\mo_2=\mo_3=\mo_4$. Therefore, the second PREE \eqref{sec2-second-PREE}
at late times is reduced to
\[
\lim\limits_{t\to\infty}e^{-\Delta S^{(2)}(\mathcal{T}^{\psi|\tilde\psi}_A(t))}=&\frac{\sum_{k\in\S}|C_k|^4L^{8h_k}\la\mo_k(w_1,\bar w_1)\mo_k^\dagger(w_2,\bar w_2)\mo_k(w_3,\bar w_3)\mo_k^\dagger(w_4,\bar w_4)\ra_{\Sigma_2}}{\left(\sum_{j\in\S}|C_j|^2L^{4h_j}\la\mo_j(w_1,\bar w_1)\mo_j^\dagger(w_2,\bar w_2)\ra_{\Sigma_1}\right)^2}\nn\\
=&\frac{\sum_{k\in\S}d_k^{-1}|C_k|^4L^{8h_k}\la\mo_k(w_1,\bar w_1)\mo_k^\dagger(w_2,\bar w_2)\ra^2_{\Sigma_1}}{\left(\sum_{j\in\S}|C_j|^2L^{4h_j}\la\mo_j(w_1,\bar w_1)\mo_j^\dagger(w_2,\bar w_2)\ra_{\Sigma_1}\right)^2},\label{n=2-replica-results}
\]
where from the first line to the second line, we utilize the known results that the four-point function of $\mo_k$ on $\Sigma_2$ at late times splits into two two-point functions on $\Sigma_1$ multiplied by the inverse of the quantum dimension of $\mo_k$ \cite{Guo:2022sfl}. Note eqn. \eqref{n=2-replica-results} perfectly matches eqn. \eqref{resultsfinalapp} for $n=2$.

The above calculation suggests that for general $n$, a $2n$-point function on $\Sigma_n$ does not vanish at late times if and only if all these $2n$ operators are the same. Indeed, in the replica calculation of $n$-th PREE,  one can readily obtain \eqref{resultsfinalapp} by simply considering $2n$-point functions involving only one primary. To show the remaining $2n$-point functions do vanish at late times, we begin with the $n=3$ case and evaluate the 6-point functions on $\Sigma_3$.\footnote{For the discussions of the higher point functions, it is more convenient to adopt the coordinate convention in \cite{He:2014mwa}.} According to the conformal map between $\Sigma_k$ and $\Sigma_1$, a general 6-point correlation function on $\Sigma_3$ can be written as
\[
&\la\mo_1(w_1,\bar w_1)\mo^\dagger_2(w_2,\bar w_2)\mo_3(w_3,\bar w_3)\mo^\dagger_4(w_4,\bar w_4)\mo(w_5,\bar w_5)\mo^\dagger_6(w_6,\bar w_6)\ra_{\Sigma_3}\nn\\
=&\prod_{i=1}^6|3 z_i^2|^{-2h_i} \la\mo_1(z_1,\bar z_1)\mo^\dagger_2(z_2,\bar z_2)\mo_3(z_3,\bar z_3)\mo^\dagger_4(z_4,\bar z_4)\mo_5(z_5,\bar z_5)\mo^\dagger_6(z_6,\bar z_6)\ra_{\Sigma_1}\label{2.1}.
\]
To evaluate the 6-point function on $\Sigma_1$, we insert a complete basis into this function
\[
&\la\mo_1(z_1,\bar z_1)\mo^\dagger_2(z_2,\bar z_2)\mo_3(z_3,\bar z_3)\mo^\dagger_4(z_4,\bar z_4)\mo_5(z_5,\bar z_5)\mo^\dagger_6(z_6,\bar z_6)\ra_{\Sigma_1}\nn\\
=&\sum\limits_{p} \lim\limits_{z_0,\bar z_0\to \infty} z_0^{2h_p}\bar z_0^{2\bar h_p}\nn\\
&\times\la\mo_1(z_1,\bar z_1)\mo^\dagger_2(z_2,\bar z_2)\mo_3(z_3,\bar z_3)\mo^\dagger_p(0,0)\ra_{\Sigma_1}\la\mo_p(z_0,\bar z_0)\mo^\dagger_4(z_4,\bar z_4)\mo_5(z_5,\bar z_5)\mo^\dagger_6(z_6,\bar z_6)\ra_{\Sigma_1}\label{2.2}
\]
By using Mobius transformation, the two 4-point functions in \eqref{2.2} can be rewritten as
\[
&\la\mo_1(z_1,\bar z_1)\mo^\dagger_2(z_2,\bar z_2)\mo_3(z_3,\bar z_3)\mo^\dagger_p(0,0)\ra_{\Sigma_1}\la\mo_p(z_0,\bar z_0)\mo^\dagger_4(z_4,\bar z_4)\mo_5(z_5,\bar z_5)\mo^\dagger_6(z_6,\bar z_6)\ra_{\Sigma_1}\nn\\
=&\left|\frac{z_{12}z_1}{z_2}\right|^{-2h_1}\left|\frac{z_{12}z_1}{z_{12}^2 z_2}\right|^{2h_2}\left|\frac{z_{12}z_1}{z_{13}^2z_2}\right|^{2h_3}\left|\frac{z_{12}z_1}{z_1^2 z_2}\right|^{2h_p
}G^{21}_{3p}(\eta,\bar \eta)\nn\\
&\times\left|\frac{z_{04}z_{06}}{z_{46}}\right|^{-2h_p}\left|\frac{z_{04}z_{06}}{z_{04}^2 z_{46}}\right|^{2h_4}\left|\frac{z_{04}z_{06}}{z_{05}^2z_{46}}\right|^{2h_5}\left|\frac{z_{04}z_{06}}{z_{06}^2 z_{46}}\right|^{2h_6
}G^{4p}_{56}(\eta',\bar \eta').\label{2.3}
\]
with $\eta=\frac{z_{12}z_3}{z_{13}z_2}$ and $\eta'=\frac{z_{04} z_{56}}{z_{05}z_{46}}$ whose late-time behavior are $(\eta,\eta')\simeq(1+\frac{x-\tilde x+2i\e}{t},1+\frac{x-\tilde x+2i\e}{t})\sim(1,1)$.

Combining \eqref{2.1}, \eqref{2.2} and \eqref{2.3}, the late-time behavior of the holomorphic part of the 6-point function on $\Sigma_3$ is
\[
&\lim\limits_{t\to \infty}\la\mo_1(w_1)\mo^\dagger_2(w_2)\mo_3(w_3)\mo^\dagger_4(w_4)\mo(w_5)\mo^\dagger_6(w_6)\ra_{\Sigma_3}\nn\\
\sim&\sum\limits_{l_1,l_2,p}\left(\prod\limits_{i=1}^6 t^{-\frac{2}{3}h_i}\right)\left(t^{-\frac{1}{3}h_1}t^{-\frac{1}{3}h_2}t^{-\frac{1}{3}h_3}t^{-\frac{1}{3}h_p}\right)t^{h_2+h_3-h_{l_1}}\left(t^{\frac{1}{3}h_p}t^{-\frac{1}{3}h_4}t^{-\frac{1}{3}h_5}t^{-\frac{1}{3}h_6}\right)t^{h_4+h_5-h_{l_2}}t^{h_1+h_6-h_p}
\]
where $h_{l_1}$ and $h_{l_2}$ are contributed from the fusion
\[
\mo_2\times\mo_3=\mo_{l_1}+\dots,\quad \mo_4\times\mo_5=\mo_{l_2}+\dots
\]
as shown in Fig. \ref{fusion1}.
\begin{figure}[h]
\centering
\includegraphics[width=0.7\linewidth]{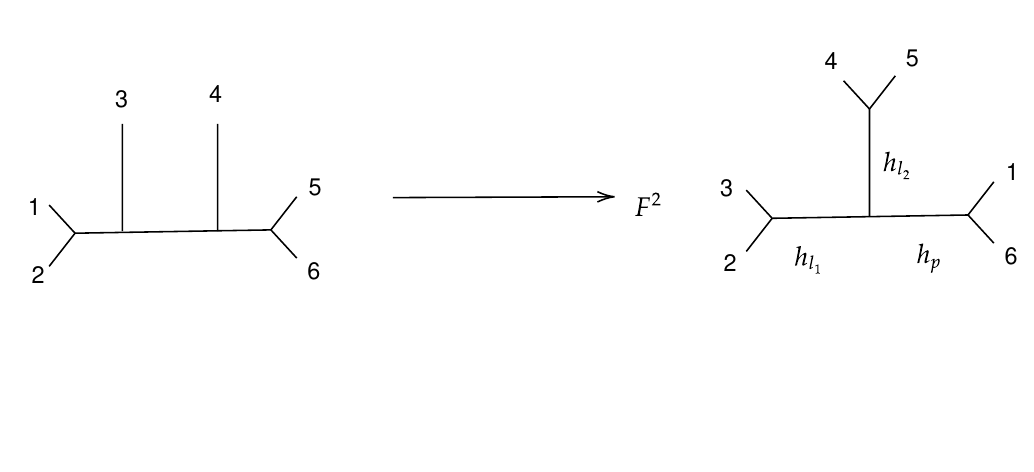}
\caption{Fusion transformations to obtain $\lim\limits_{t\to\infty}\la\mo_1(w_1)\mo^\dagger_2(w_2)\mo_3(w_3)\mo^\dagger_4(w_4)\mo(w_5)\mo^\dagger_6(w_6)\ra_{\Sigma_3}$}\label{fusion1}
\end{figure}

Using the result, we derive for the 4-point function \eqref{four-fusion}, the 6-point function on $\Sigma_3$ does not vanish at late times if, and only if
\[
h_{l_1}=h_{l_2}=h_p=0
\]
yielding the fusion rules
\[
\mo_2\times\mo_3=\mathbb{I}+\dots,\quad \mo_4\times\mo_5=\mathbb{I}+\dots,\quad\mo_1\times\mo_6=\mathbb{I}+\dots.
\]
Similarly,  the late-time behavior of the anti-holomorphic part of the 6-point function yields the fusion rules
\[
\mo_1\times\mo_2=\mathbb{I}+\dots,\quad \mo_3\times\mo_4=\mathbb{I}+\dots,\quad\mo_5\times\mo_6=\mathbb{I}+\dots,
\]
which means that the 6-point function on $\Sigma_3$ does not vanish at late times only when
\[
\mo_1=\mo_2=\mo_3=\mo_4=\mo_5=\mo_6.
\]
All the discussion of the 6-point function on $\Sigma_3$ can be extended to a general $2n$-point function on $\Sigma_n$ by inserting a complete basis and using the conformal block, and the results are similar, implying that the $2n$-point function does not vanish if and only if all the operators in the $2n$-point function belong to the same conformal family.
 \subsection{Pseudo entropy amplification}
{In this subsection, we introduce the phenomenon of pseudo entropy amplification, where PE (PREE) contains all the information about the linear combination operator whereas EE (REE) does not, through the evaluation of the maximal value of \eqref{resultsfinalapp} by adjusting the combination coefficients $C_k$.}
\subsubsection{The REE and EE}\label{EEandREE}
 Let us begin by focusing on the REE and EE. By simply removing the {tildes} in \eqref{resultsfinalapp}, we arrive at the following result
\[
\lim_{t\to\infty}\Delta S^{(n)}\big(\rho_A^{\psi}(t)\big)=&\lim_{t\to\infty}\Delta S^{(n)}\big(\T_A^{\psi|\psi}(t)\big)\nn\\
=&\frac{1}{1-n}\log\frac{\sum_{k\in\S}d_k^{1-n}|C_k|^{2n}(\frac{L}{2\epsilon})^{4nh_k}}{(\sum_{j\in\S}|C_{j}|^2(\frac{L}{2\epsilon})^{4h_{j}})^n}\nonumber\\
%=&\lim_{\epsilon\to0}\frac{1}{1-n}\log\frac{\sum_pd_p^{1-n}|C_p|^{2n}(2\epsilon)^{4n(h_m-h_p)}}{\big(\sum_{p'}|C_{p'}|^2(2\epsilon)^{4(h_m-h_{p'})}\big)^n}\nonumber\\
=&\frac{1}{1-n}\log\frac{\sum_{k\in S_M}d_k^{1-n}|C_k|^{2n}}{(\sum_{j\in\mathcal{S}_{M}}|C_{j}|^2)^n},\label{REE-for-heavy-operators}
\]
where $\S_M:=\big\{\mo_k|\mo_k\in\S, h_k=h_M\equiv\max_{l\in\S}\{h_l\}\big\}$. From the second equation to the third equation, we omit terms associated with lighter operators since, in the limit as $\e\to0$, they constitute sub-leading contributions within the sum. Consequently, we conclude that information about lighter operators is lost in the REE when $\S_M$ is a proper subset of $\S$. This conclusion holds, especially in cases where a single heaviest primary operator with a quantum dimension of $d_M$ exists, simplifying eqn. \eqref{REE-for-heavy-operators}   to  $\log d_M$.
For cases where the conformal dimensions of primary operators in \eqref{simplified-linear-combination-operator} are the same\footnote{We continue to assume that the primary operators are orthogonal to each other for this case.}, the late-time behavior of $n$-th REE  for $\rho_A^{\psi}$ is
\[
\lim_{t\to\infty}\Delta S^{(n)}(\rho_A^{\psi}(t))=\frac{1}{1-n}
\log\left(\sum_{k\in\S}d_k^{1-n}|C_k|^{2n} \right),
\]
where the constraint $\sum_{k\in\S} |C_k|^2=1$ was employed to simplify the expression.

Taking the limit of $n\to 1$ to eqn. \eqref{REE-for-heavy-operators}, we obtain the expression of the late-time EE,
\[
\lim_{t\to\infty}\Delta S\big(\rho_A^{\psi}(t)\big)=\sum_{k\in\S_M}p_k\log d_k+H(p_k), \label{EE-for-heavy-operators}
\]
where $p_k=\frac{|C_k|^2}{\sum_{j\in S_M}|C_j|^2}$ is an effective probability distribution, and $H(p_k)=-\sum_{k\in\S_M}p_k\log p_k$ the classical Shannon entropy of $\{p_k\}$. %{\color{red}Note that eqn. \eqref{EE-for-heavy-operators} is reduced to eqn. \eqref{EE-of-rhoA}  when $\S_M=\S$.}

As functions of the combination coefficients $C_k$, the maximum values of \eqref{REE-for-heavy-operators} and \eqref{EE-for-heavy-operators} can be determined using \textit{H\"{o}lder's inequality}, a generalization of \textit{Cauchy-Schwarz~inequality},
\[
\frac{1}{n-1}\log\frac{\left(\sum_{k\in\S_M}d_k^{1-\frac{1}{n}}(y_k^{n})^{\frac{1}{n}}\right)^n}{\sum_{j\in\S_M}y_k^n}\leq& \frac{1}{n-1}\log\frac{\left((\sum_{k\in\S_M} d_k)^{\frac{n-1}{n}}(\sum_{k\in\S_M} y_k^n)^\frac{1}{n}\right)^n}{\sum_{j=1}^{m}y_j^n}\label{EE-Holder}\\
=&\log\left(\sum_{k\in\S_M}d_k\right).\nn
\]
 Here, we
redefine a series of parameters in terms of $C_k$
\[
y_k:= |C_k|^2d_k^{\frac{1}{n}-1}>0,\quad(k=1,2,...,m).
\]
The equality in \eqref{EE-Holder} holds if and only if
\[
y_k^n\propto d_k\Rightarrow |C_k|\propto\sqrt{d_k}\label{EE-propotion}.
\]
Thus, we conclude that the maximum late-time REE and EE saturates to the logarithm of the sum of quantum dimensions of the heaviest operators,
\[
\max_{\{C_k\}}\big\{\lim\limits_{t\to\infty}\Delta S^{(n)}\big(\rho_A^{\psi}(t)\big)\big\}=\max_{\{C_k\}}\big\{\lim\limits_{t\to\infty}\Delta S\big(\rho_A^{\psi}(t)\big)\big\}=\log\left(\sum_{j\in\S_M}d_j\right).\label{maximum-REE-section2}
\]
The corresponding combination coefficients, denoted as $C_k^\star$, are given by
\[
 |C_k^\star|=\sqrt{\frac{d_k}{\sum_{j\in\S}d_j}}.\label{cstar-EE-section2}
\]
%where we fix the redundancy of $\{C_k\}$ by imposing the normalization constraint $\sum_{k\in\S}|C_k|^2=1$.
Note that there remains an undetermined degree of phase freedom for each coefficient $C_k$ since the late-time formula \eqref{resultsfinalapp} depends only on the magnitude of $C_{k}$.

We can examine our conclusions in specific RCFT models.
One typical example in RCFT is the linear combination of vertex operators in 2D massless free scalar theory.\footnote{We express our gratitude to Tadashi Takayanagi for bringing this to our attention.} As demonstrated in \cite{Caputa:2015tua,Bhattacharyya:2019ifi}, the variation of the second REE of the one-parameter primary operator $V_q=\sqrt{q}e^{i\a\phi}\pm\sqrt{1-q}e^{-i\a\phi}$ is maximized when the parameter $q$ equals $\frac{1}{2}$,
\[
\max_q\{\Delta S_A^{(2)}(V_q)\}=\Delta S_A^{(2)}(V_\frac{1}{2})=\log 2=\log (1+1),
\]
 where $1$ represents the quantum dimension of the vertex operators $e^{\pm i\a\phi}$ and $q=\frac{1}{2}$ maximizing $\Delta S_A^{(2)}(V_a)$ precisely aligns with the obtained result.
%As mentioned in section \ref{section 1:intro},  when $\S_M=\S$, i.e., the conformal dimensions of primary operators in \eqref{simplified-linear-combination-operator} are the same, the above results (eqn. \eqref{REE-for-heavy-operators}, \eqref{EE-for-heavy-operators}, and \eqref{cstar-EE-section2}) perfectly match the computations in free scalar in \cite{Caputa:2015tua}. The operators' information is not lost in this case.
A more nontrivial verification can be made in the critical Ising model. Our results indicate that the information of lighter operators in the critical Ising model will be lost in REE/EE in the limit of $\e\to0$, as the conformal dimensions of the three primary operators (i.e., the Ising spin $\sigma$, energy density $\varepsilon$, and identity $\mathbb{I}$) in the critical Ising model differ from each other.

We numerically compute the late-time second REE for two linear combination operators: $C_\sigma\cdot L^{2h_{\sigma}} \sigma+C_{\varepsilon}\cdot L^{2h_{\epsilon}}\varepsilon$  and  $C_\sigma\cdot L^{2h_{\sigma}} \sigma+C_{\mathbb{I}}\cdot \mathbb{I}$ (See Fig. \ref{Fig1:un-normal-linear-combination}(a) and (b) respectively). Our late-time formula \eqref{resultsfinalapp} (solid lines) exhibits good agreement with the numerical data (hollow circles). For the linear combination of the Ising spin and the energy density (Fig. \ref{Fig1:un-normal-linear-combination}(a)), the late-time second REE approaches zero in the  $\epsilon\to0$ limit, as the heavier operator $\varepsilon$ possesses a quantum dimension of $1$. Conversely, for the linear combination of the Ising spin and the identity (Fig. \ref{Fig1:un-normal-linear-combination}(b)), the late-time second REE approaches $\log\sqrt{2}$ in the  $\epsilon\to 0$ limit since the heavier operator $\sigma$ has a quantum dimension of $\sqrt{2}$.

\begin{figure}[t]
	\centering
\captionsetup[subfloat]{farskip=5pt,captionskip=1pt}
\subfloat{
			\includegraphics[width =0.45\linewidth]{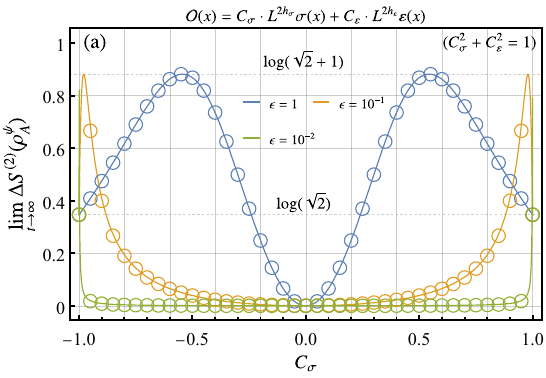}}
\hfill
\subfloat{
			\includegraphics[width =0.45\linewidth]{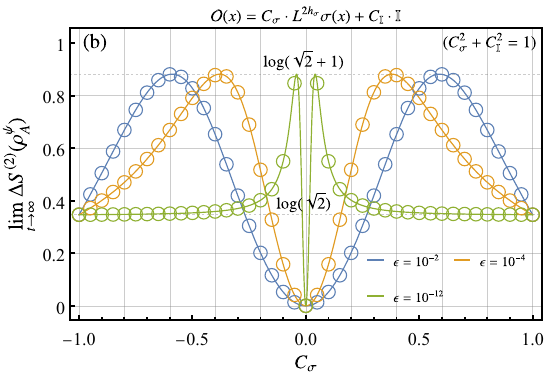}}\\
   \subfloat{
			\includegraphics[width =0.45\linewidth]{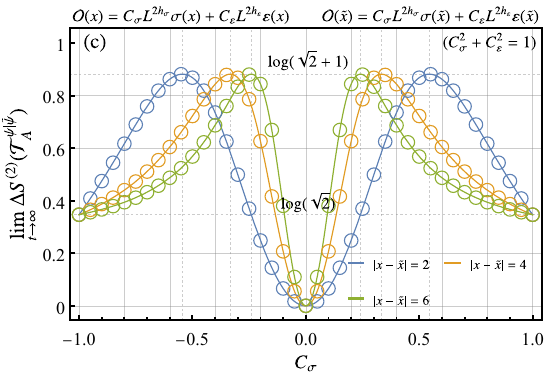}}
   \hfill
    \subfloat{
			\includegraphics[width =0.45\linewidth]{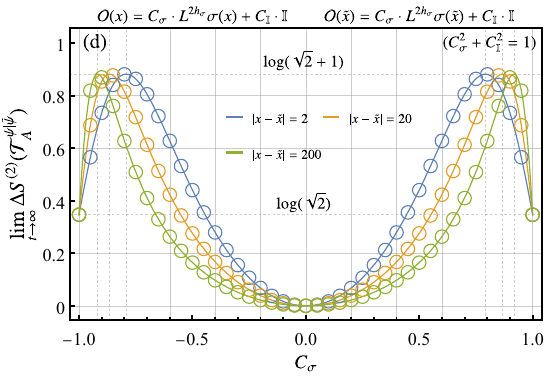}}
	\caption{The variation of the late-time second REE (in panels (a) and (b)) and the late-time second PREE (in panels (c) and (d)) with respect to the combination coefficient $C_{\sigma}$ in the critical Ising model. The solid lines are plotted using the late-time formula \eqref{resultsfinalapp}, while the hollow circles denote numerical data computed using the well-known four-point functions in the critical Ising model.  The vertical dotted lines in panels (c) and (d), plotted in terms of eqn. \eqref{Cstar-secton2-PREE}, correspond to the maximum values of the second PREE. All four panels are symmetric along $C_\sigma=0$ since the late-time formula \eqref{resultsfinalapp} depends solely on the magnitude of the combination coefficients. Note that the UV regulator $\e$, while a sensitive parameter in REE calculations, does not impact PREE computations and can be safely set to 0. }
	\label{Fig1:un-normal-linear-combination}
	\vspace{-0.5em}
\end{figure}

\subsubsection{The PREE and PE}
After exploring the late-time behavior and the maximum values of REE and EE in the previous subsection, we now focus on studying PREE and PE. This part demonstrates how the PREE and PE can preserve information about all lighter operators in $\S$. This preservation is achieved by simply shifting the excitation position of $\mo$ from $x$ to $\tilde{x}$ to create the final state $|\tilde\psi\ra$ \eqref{psi-tilde}.

The finiteness of the difference between $x$ and $\tilde x$ prevents the divergence of two-point functions in eqn. \eqref{resultsfinalapp} and leaves a new expression for the late-time PREE in the $\e\to0$ limit,
\[
\lim_{t\to\infty}\Delta S^{(n)}(\T^{\psi|\tilde\psi}_A(t))=&\frac{1}{1-n}\log\left[\sum_{k\in\S}\left(\frac{d_k^{\frac{1}{n}-1}|C_k|^2\big(\frac{L}{|x-\tilde x|}\big)^{4h_{k}}}{\sum_{j\in\S}|C_{j}|^2\big(\frac{L}{|x-\tilde x|}\big)^{4h_{j}}}\right)^n\right].\label{nth-PREE-explict-form}
\]
Since the summation is performed on all operators in $\S$, information about lighter operators will be encoded in the PREE. Compared with the results of EE, we observe the phenomenon of pseudo entropy amplification \cite{Ishiyama:2022odv}. It can be shown more explicitly by looking at the maximum value of the $n$-th PREE. Let us define a series of parameters $y_k$, like before,
\[
y_k:=d_k^{\frac{1}{n}-1}|C_k|^2\big(\frac{L}{x-\tilde x}\big)^{4h_k}.
\]
The $n$-th PREE \eqref{nth-PREE-explict-form} is then transformed into the same expression in \eqref{EE-Holder}. The H\"{o}lder's inequality guarantees that \eqref{nth-PREE-explict-form} reaches its maximum value of $\log\big(\sum_{k\in\S} d_k\big)$ when the combination coefficients are given by
\[
|C^\star_k|=\sqrt{\frac{\frac{d_k}{L^{4h_k}\la\mo_k^\dagger(\tilde x)\mo_k(x)\ra}}{\sum_{j\in\S}\frac{d_j}{L^{4h_j}\la\mo_j^\dagger(\tilde x)\mo_j(x)\ra}}}.\label{Cstar-secton2-PREE}
\]
The maximum value of the late-time PREE encompasses the quantum dimension of each primary operator within the set $\S$. This outcome contrasts with the result of REE \eqref{maximum-REE-section2}, thereby explicitly reflecting the pseudo entropy amplification phenomenon.  On the other hand, the late-time PE of $\mathcal{T}_A^{\psi|\tilde\psi}(t)$, obtained by taking the limit as $n\to 1$ for \eqref{nth-PREE-explict-form}, is given by
\[
\lim_{t\to\infty}\Delta S\big(\mathcal{T}_A^{\psi|\tilde\psi}(t)\big)=H\big(p_k(x,\tilde x)\big)+\sum_{k\in\S} p_k(x,\tilde x)\log d_k.\label{PEE_final}
\]
In the above, $p_k(x,\tilde x):=\frac{|C_k^2|L^{4h_k}\la\mo_k^\dagger(\tilde x)\mo_k(x)\ra}{\sum_{j\in\S}|C_j^2|L^{4h_j}\la\mo_j^\dagger(\tilde x)\mo_j(x)\ra}$ is a position-depended probability distribution.

Similarly to the REE, we can confirm the accuracy of our results using numerical computations in the critical Ising model. As illustrated in Fig. \ref{Fig1:un-normal-linear-combination}(c) and Fig. \ref{Fig1:un-normal-linear-combination}(d), we numerically calculate the second PREE for linear combination operators $\mo(x)$ and $\mo(\tilde x)$, where $\mo$ represents a linear combination of the Ising spin and the energy density or the Ising spin and the identity. Our late-time formula \eqref{resultsfinalapp} (solid lines)  consistently matches the numerical data (hollow circles).
Furthermore, the coefficients corresponding to the maximum of the late-time 2nd PREE are also accurately characterized by eqn. \eqref{Cstar-secton2-PREE}. By comparing the green line in Fig. \ref{Fig1:un-normal-linear-combination}(a) with the lines in Fig. \ref{Fig1:un-normal-linear-combination}(d), one can readily observe the phenomenon of pseudo entropy amplification.

\subsection{A connection between Fusion numbers and OPE coefficients}\label{section:fusion and OPE}
The results in the above compellingly demonstrate the efficacy and simplicity of the Schmidt decomposition method in analyzing the late-time EE (PE) of linear combinations of operators, eliminating the need for computing intricate higher-point functions. This section will mainly utilize the Schmidt decomposition method to explore the relationship between OPE coefficients and fusion coefficients.

Starting from the OPE \eqref{OPE}, one can write down an equality for the locally excited states,
\[
e^{-(\e+it) H}\mo_1(x+1)\mo_2(x)|\Omega\ra=\sum\limits_k \sum\limits_{\{\alpha,\bar \alpha\}} C_{k12}\beta_{12}^{k\{\alpha\}}\bar{\beta}^{k\{\bar\a\}}_{12}e^{-(\e+it) H}\mo_k^{\{\alpha,\bar \alpha\}}(x)|\Omega\ra.\label{OPE-state}
\]
As shown in \cite{Guo:2018lqq}, the EE of the state on the l.h.s. of \eqref{OPE-state} saturates to $\log (d_1d_2)$ at late times. \footnote{One might consider substituting $x+1$ with $x+2\epsilon$ to yield the second type of linear combination operators \eqref{normalized-linear-combination-operator} on the r.h.s. of the OPE. However, it's essential to note that the EE of the composite operator on the l.h.s. of the OPE, after the substitution, does not converge to $\log(d_1d_2)$ as $\epsilon$ tends to zero and $t$ tends to infinity \cite{Guo:2018lqq}.} To calculate the EE of the  state on the r.h.s. of \eqref{OPE-state},  {building on the Schmidt decomposition method, we would like to recast the state  into the following form}
\[
|\psi_{12}(t)\ra=\frac{1}{\sqrt{\sum_{k}|\tilde C_{k12}(\e)|^2}}\sum_{k}\tilde{C}_{k12}(\e)\cdot|\psi_{k12}(t)\rangle,
\]
with
\[
|\psi_{k12}(t)\ra:=&\frac{1}{\sqrt{\N_{k12}(\e)}}\sum_{\{\a,\bar\a\}}\beta_{12}^{k\{\a\}}\bar{\b}_{12}^{k\{\bar\a\}}e^{-(\e+it)H}\mo_k^{\{\a,\bar\a\}}(x)|\Omega\ra,\quad\big(\la\psi_{k12}(t)|\psi_{l12}(t)\ra=\d_{kl}\big),\label{psi12}\\
\N_{k12}(\epsilon):=&\sum_{\{\a,\bar\a,\g,\bar\g\}}|{2\e}|^{-(4h_k+|\a|+|\bar\a|+|\gamma|+|\bar\gamma|)}e^{\frac{i\pi}{2}(|\bar\a|+|\bar \g|-|\a|-  |\g|)}\b^{k\{\a\}}_{12}\b^{k\{\g\}}_{12}\bar\b^{k\{\bar\a\}}_{12}\bar\b^{k\{\bar\g\}}_{12}c_k(\{\a,\bar\a\},\{\g,\bar\g\}),
\]
and
\[
\tilde C_{k12}(\e):=C_{k12}\cdot\sqrt{\N_{k12}(\e)}.
\]
In the above, we denote by $c_k(\{\a,\bar\a\},\{\g,\bar\g\})$ the coefficient of the two-point function\\ $\la\mo^{\{\a,\bar\a\}}_k(z,\bar z)\mo^{\{\g,\bar\g\}}_k(0,0)\ra$ (e.g., $c_k(\{-1,0\},\{0,0\})=-2h_k$).
Since $|\psi_{k12}(t)\rangle$ are orthogonal to one another, we can apply the Schmidt decomposition method like before to derive the EE of $|\psi_{12}\ra$ at late times, which is given by
\[
\lim\limits_{t\to\infty}\Delta S\big(\rho_A^{\psi_{12}}(t)\big)=H(p_{k})+\sum_{k}p_k\cdot\lim_{t\to\infty}\Delta S\big(\rho^{\psi_{k12}}_A(t)\big),
\]
where $p_k\equiv{|\tilde C_{k12}|^2}/{\sum_j|\tilde C_{j12}|^2}$. Due to the non-orthogonality among distinct locally descendant excited states in eqn. \eqref{psi12}, it's unattainable to obtain the EE of $\rho_A^{\psi_{k12}}$ via Schmidt decomposition method. Nevertheless, we can still compute it by employing the replica method. As demonstrated in \cite{Chen:2015usa,Guo:2022sfl}, for a general linear combination of descendant operators (within the same conformal family, say $[\mo_k]$), there exists an additional correction to the subsystem's EE at late times apart from $\log d_k$ (for instance, $\lim_{t\to\infty}\Delta S_A(\partial \mo_k+\bar{\partial}\mo_k)=\log d_k+\log 2$). However, as shown in \cite{Guo:2018lqq}, due to the fact that the linear combination operator on the r.h.s. of the OPE need to match the composite operators on the l.h.s. under the same conformal mapping in RCFTs, this additional correction was found to be zero for the state $|\psi_{k12}\ra$ \eqref{psi12}. Consequently, we have
\[
\lim\limits_{t\to\infty}\Delta S\big(\rho_A^{\psi_{12}}(t)\big)=H(p_{k})+\sum_{k}p_k\cdot\log d_k=\log\sum_k d_k,\label{e78}
\]
where the second equality arises from the EE of the composite operator on the l.h.s. of the OPE and the fusion algebra \eqref{fusion-quantum-dimension}. {On the other hand, as per the conclusions from previous sections, the second equality ensures that $p_k$ will maximize the expression in eqn. \eqref{e78}.} We thus arrive at an equality between the quantum dimensions and OPE coefficients
\[
p_k=\frac{d_k}{\sum_j d_j}=\frac{|\tilde C_{k12}|^2}{\sum_j|\tilde C_{j12}|^2}\Rightarrow\sqrt{\frac{d_k}{d_l}}=\frac{C_{k12}}{C_{l12}}\lim_{\e\to0}\sqrt{\frac{|\mathcal{N}_{k12}(\e)|}{|\mathcal{N}_{l12}(\e)|}}.\label{eqn77}
\]
{Notice that the r.h.s. of the above equality contains information about the operators $\mo_1$ and $\mo_2$, while the l.h.s. does not. Hence the above equality is not complete. One may replace\footnote{{Such a replacement may be understood in terms of the following way: we recast eqn. \eqref{e78} into the following form to reveal the dependence of the result on the fusion coefficients,
\[
\lim\limits_{t\to\infty}\Delta S\big(\rho_A^{\psi_{12}}(t)\big)=H(p_{k})+\sum_{k}p_k\cdot\log N^{k}_{12}d_k=\log\sum_k N_{12}^kd_k,\quad N^{k}_{12}\in\{0,1\}.\label{eq2311}
\]Note that $p_k=0$ if $N_{12}^k=0$ and we define $0\log0=0$.
Similar to eqn. \eqref{eqn77}, eqn. \eqref{eq2311} holds if and only if $p_k=\frac{N_{12}^kd_k}{\sum_jN_{12}^jd_j}$.
}} $d_{k(l)}$ on the l.h.s. of eqn. \eqref{eqn77} with $N^{k(l)}_{12}d_{k(l)}$, yielding an equality relating fusion coefficients to OPE coefficients, to restore the information of $\mo_1$ and $\mo_2$ for the l.h.s. of eqn. \eqref{eqn77},}
\[
\sqrt{\frac{N_{12}^kd_k}{N_{12}^ld_l}}=\frac{C_{k12}}{C_{l12}}\lim_{\e\to0}\sqrt{\frac{|\mathcal{N}_{k12}(\e)|}{|\mathcal{N}_{l12}(\e)|}},\quad N^{k}_{12}\in\{0,1\}. \label{formula:FusionAndOPE}
\]
While we have linked fusion coefficients and quantum dimensions to coefficients of OPE through EE, rigorously verifying the above two equations necessitates dealing with the infinite summation over secondary fields. Even in the simplest Ising model, this remains a rather challenging task. In this regard, we aim to seek relevant insights in future work.

\section{The entanglement dynamics of refined linear combination operators \eqref{normalized-linear-combination-operator}}\label{section 3 discussion}
%Up to this point, we have deliberated on the EE and PE of two types of linear combination operators (type I: eqn. \eqref{simplified-linear-combination-operator}, type II: eqn. \eqref{normalized-linear-combination-operator}). The findings suggest that, when discussing EE, the second type of operator appears to be superior to the first, as EE fails to extract information about lighter operators within the first type.  In this section, we will comment further on the EE concerning both types of operators. We shall see that when attempting to establish a connection between OPE coefficients and fusion numbers using the entanglement of local operators, the type I operators may be more suitable than the type II. Additionally, we will utilize the reduced density matrices generated by the type II operators to illustrate the late-time decoherence-like phenomenon.
In the previous section, we provided a connection between OPE coefficients and fusion coefficients by examining the EE of a linear combination of operators on the r.h.s. of the OPE. Such a linear combination operator is distinctive as it involves linear combinations of an infinite number of descendant operators. When we confine the number of operators to a finite set, as in eqn. \eqref{simplified-linear-combination-operator}, we encounter an intriguing phenomenon, i.e., the REE/EE, when applied to a general linear combination operator $\mo$ \eqref{simplified-linear-combination-operator}, has limitations in capturing information about lighter operators. In this section, we address this limitation\footnote{We would like to thank Yuya Kusuki for valuable discussions regarding this section.} by redefining the linear combination operator $\mo$ \eqref{simplified-linear-combination-operator} as the form of \eqref{normalized-linear-combination-operator}. We then show more details about the entanglement dynamics of such refined operators.

\subsection{Recovering information of lighter operators in the EE}\label{section2.3}
To understand the effectiveness of this redefinition, we begin by elucidating why the REE and EE of linear combination operators in \eqref{simplified-linear-combination-operator}  fail to capture information about lighter operators. From the perspective of replica calculations, this arises due to the varying divergence of the $2n$-point functions on the numerator at late times as $\epsilon$ approaches zero. Despite $L$ canceling out the dimension of $\epsilon$, allowing for the summation of correlation functions associated with different operators, only the most divergent portion is ultimately retained. While this explanation is accurate, it lacks intuitiveness because, from the definition of the linear operator in \eqref{simplified-linear-combination-operator}, all primary operators seem to be equally weighted. Approaching the issue from the perspective of quantum states rather than operators provides a more satisfactory answer. When discussing the EE of locally excited states, we should focus on normalized quantum states rather than operators. Consequently, we prefer to regard $|\psi\ra$ \eqref{simplified-linear-combination-operator} as a linear combination of normalized locally primary excited states in the Hilbert space rather than a state generated by a linear combination operator, i.e., we have
\[
|\psi(t)\rangle=&\frac{1}{\sqrt{\mathcal{N}(\e)}}\sum_{k\in\S}C_kL^{2h_k}e^{-iH(t-i\e) }\mo_k(x)|\Omega\rangle\nn\\
=&\frac{1}{\sqrt{\mathcal{N}(\e)}}\sum_{k\in\S} C_k\cdot\Big(\frac{L}{2\epsilon}\Big)^{2h_k}|\mo_k(x,t)\rangle,\quad(\la\psi(t)|\psi(t)\ra=1),\label{state1}
\]
where the state $|O_k(x,t)\ra$, as defined in \eqref{Opstate}, carries the information of entanglement among subsystems when the overall system is quenched by $\mo_k(x)$ at the initial moment.
The above expression indicates that, from the perspective of quantum states, the action of the operator $\mo(x)$ \eqref{simplified-linear-combination-operator} on the vacuum results in a superposition of heavy operator states, with the weight of light operator states approaching zero in the limit as $\epsilon$ tends to zero\footnote{$\S_M$ in \eqref{epsi0} denotes the set consisting of the heaviest operators within the set $\S$, as introduced in section\ref{section 2.1}.},
\[
\lim_{\epsilon\to0}|\psi(t)\rangle=\frac{1}{\sqrt{\sum_{k\in\S_M}|C_k|^2}}\sum_{k\in\S_M}C_k\cdot|\mo_k(x,t)\rangle.\label{epsi0}
\]

{The method to restore the contributions of light primary operators is currently evident: by following the expression in eqn. \eqref{state1}, and by substituting $L$ with $2\epsilon$ (i.e., considering linear combination operators taking the form of \eqref{normalized-linear-combination-operator}), we can extract information about all operators from the EE.} The state generated by the refined linear combination operator \eqref{normalized-linear-combination-operator} is a superposition of all locally primary excited states,
 \[
|\psi(t)\ra=\sum_{k\in\S}C_k\cdot|\mo_k(x,t)\ra,\quad(\sum_{k\in\S}|C_k|^2=1).\label{3.1states}
 \]
 Utilizing two analytical methods introduced before, the $n$-th REE and EE of the above state are given by
 \[
\lim_{t\to\infty}\Delta S^{(n)}\big(\rho^{\psi}_A(t)\big)=&\frac{1}{1-n}\log\left(\sum_{k\in\S}d_k^{1-n}|C_k|^{2n}\right),\label{normal-latetime}\\
\lim_{t\to\infty}\Delta S\big(\rho^{\psi}_A(t)\big)=&H(p_k)+\sum_{k\in\S}p_k\log d_k,
 \]
 where $p_k\equiv |C_k|^2$. Both REE and EE reach their maximum value of $\log(\sum_{k\in\S}d_k)$ when the combination coefficients are given by $|C_k|=|C_k^\star|\equiv\sqrt{\frac{d_k}{\sum_{j\in\S}d_j}}$.
 We can also validate our results using numerical replica computations in the critical Ising model, as illustrated in Fig. \ref{Fig3:normal}.

\begin{figure}[t]
	\centering
\captionsetup[subfloat]{farskip=5pt,captionskip=1pt}
\subfloat{
			\includegraphics[width =0.45\linewidth]{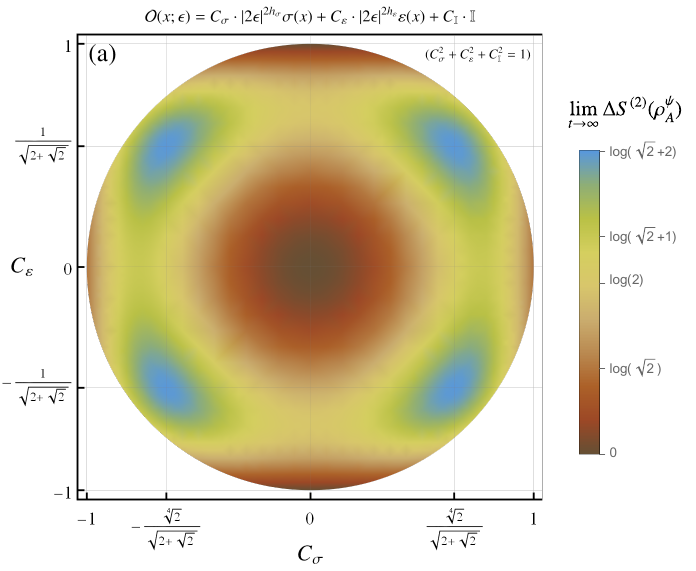}}
\hfill
\subfloat{
			\includegraphics[width =0.45\linewidth]{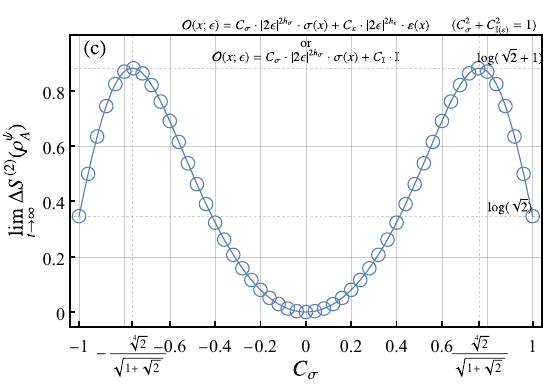}}\\
   \subfloat{
			\includegraphics[width =0.45\linewidth]{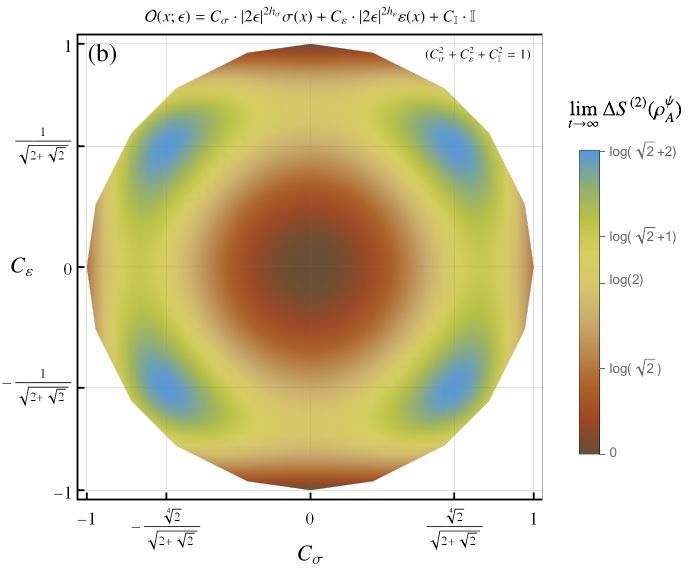}}
   \hfill
    \subfloat{
			\includegraphics[width =0.45\linewidth]{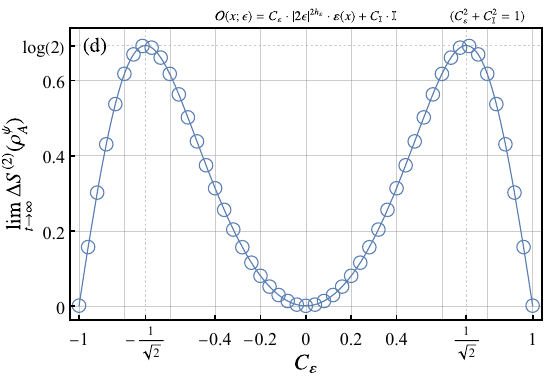}}
	\caption{The variation of the late-time second REE  with respect to the combination coefficients in the critical Ising model. Panels (a) and (b): A mixture of spin $\sigma$, energy density $\varepsilon$, and identity $\mathbb{I}$. Panels (c) and (d) correspond to linear combinations of $\varepsilon$ with $\sigma$ and $\varepsilon$ with $\mathbb{I}$, respectively. Panel (b) as well as the hollow circles in panels (c) and (d) represent numerical data obtained from known correlation functions in the critical Ising model. Panel (a) as well as the solid lines in panels (c) and (d) are plotted according to eqn. \eqref{normal-latetime}. Note that the maximum value of the second REE is $\log(\sqrt{2}+2)$, as the quantum dimensions of three primary are $d_\sigma=\sqrt{2}$, $d_{\varepsilon}=d_{\mathbb{I}}=1$.}
	\label{Fig3:normal}
	\vspace{-0.5em}
\end{figure}

\subsection{The quasiparticle picture}
 As mentioned in the introduction, for integrable theories such as RCFTs, each time-evolving locally primary or descendant excited state \eqref{Opstate} possesses an interpretation of the classical picture of quasiparticles propagation \cite{Nozaki:2014hna,He:2014mwa,Caputa:2015tua,Chen:2015usa,Numasawa:2016kmo}: at time $t=0$, a pair of entanglement quasiparticles excited by the operator $\mo_k$ emerge at position $x$, and subsequently propagate at the speed of light to positive and negative infinity respectively (although, with spatially inhomogeneous Hamiltonians such as Hamiltonians under sine-square deformation \cite{Gendiar:2008udd,Katsura:2011zyx}, the quasiparticles no longer travel at the speed of light\cite{Wen:2018vux,Goto:2021sqx,Nozaki:2023fkx}). Naturally, one would inquire whether the classical picture of quasiparticles propagation still holds for linear combinations of these locally excited states. Although it has been observed that for combinations of primary operators with the same scaling dimensions, the quasiparticles picture remains applicable \cite{Caputa:2015tua,Bhattacharyya:2019ifi,Zhang:2019kwu}. It's still worrisome that this classical picture may fail in general superposition states. On one hand, a combination of primary operators with differing scaling dimensions cannot be regarded as a primary operator anymore. On the other hand, within quantum mechanics, we know that superpositions of coherent (or quasiclassical) states generally no longer possess classical correspondences. %This parallels the idea that in the AdS/CFT context, linear combinations of geometric states might not have a gravity dual anymore \cite{Almheiri:2016blp,Guo:2018fnv}.
We can {check} the validity of the classical picture by analyzing the full-time evolution of the EE of superposition states. If the EE evolution of a certain superposition state exhibits a step-like pattern (e.g., see Fig. \ref{Fig4:quasiparticle-pic-or-break}), we infer that the classical interpretation of quasiparticles propagation holds for that state. Otherwise, we do not consider it to have a classical interpretation.
\begin{figure}[h]
    \centering
    \includegraphics[width =0.45\linewidth]{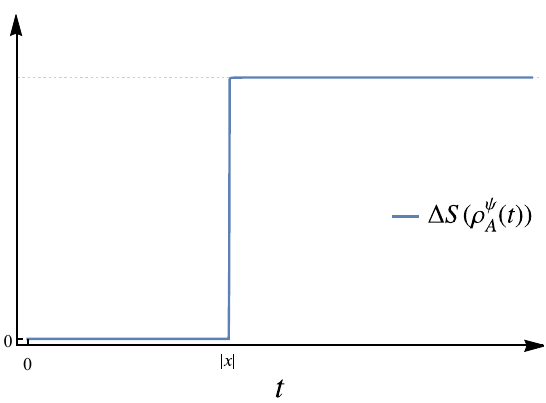}
    \caption{The complete time evolution of EE,  which exhibits a step-like behavior.}
    \label{Fig4:quasiparticle-pic-or-break}
\end{figure}

Let us reanalyze the REE of a state $|\psi\ra$ generated by the refined linear combination operator
\[
\mo(x;\e)=&\sum_{k\in\S}C_k\cdot (2\e)^{2h_k}{\mo_k(x)},\nn\\
\big(\sum_{k\in\S}|C_k|^2=1,\quad|\psi(t)\ra\equiv& e^{-iHt}|\psi\ra=\frac{1}{\sqrt{\mathcal{N}(\e)}}e^{-(\e+it) H}\mo(x,\e)|\Omega\ra\big)\label{psiyiban}
\]
using the replica method. We keep subsystem $A$ as $[0,\infty)$ and mainly focus on the case of $n=2$ for simplicity since extending this proposition to any $n$ is straightforward. As shown in section \ref{section: replica method}, utilizing the replica method, the
second REE is reduced to a series of four-point functions on a 2-sheeted Riemann surface $\Sigma_2$
\[
e^{-\Delta S^{(2)}\big(\rho^{\psi}_A(t)\big)}=\sum_{ijkl\in\S}C_iC^*_jC_kC_l^*(2\e)^{2(h_i+h_j+h_k+h_l)}\la\mo_i(w_1,\bar w_1)\mo^\dagger_j(w_2,\bar w_2)\mo_k(w_3,\bar w_3)\mo^\dagger_l(w_4,\bar w_4)\ra_{\Sigma_2}.\label{2eesec3}
\]
The four-point functions in the sum, with the help of the conformal mapping $w=z^2$, can be expressed as four-point functions on the plane $\Sigma_1$ (we take $\{i,j,k,l\}=\{1,2,3,4\}$ for example),
\[
\la\mo_1(w_1,\bar w_1)\mo^\dagger_2(w_2,\bar w_2)\mo_3(w_3,\bar w_3)\mo^\dagger_4(w_4,\bar w_4)\ra_{\Sigma_2}=&\left(\bigg|\frac{2z_1z_{12}z_{14}}{z_{24}}\bigg|^{-2h_1}\prod_{i=2}^4\bigg|\frac{2z_iz^2_{1i}z_{24}}{z_{12}z_{14}}\bigg|^{-2h_i}\right)G^{21}_{34}(\eta,\bar\eta),\label{4pt-Sigma2->Sigma1}
\]
where $(z_i,\bar z_i)$ are defined in eqn. \eqref{z1234} and $(\eta,\bar\eta)\equiv(\frac{z_{12}z_{34}}{z_{13}z_{24}},\frac{\bar z_{12}\bar z_{34}}{\bar z_{13}\bar z_{24}})$ is cross ratio. $G^{21}_{34}(\eta,\bar\eta)$ can be expanded as eqn. \eqref{conformal-block-expansion} in terms of conformal blocks $F^{21}_{34}(p|\eta)$.

At early times $0<t<|x|$, we find that the cross ratio becomes
\[
\eta\simeq\frac{\e^2}{4(|x|-t)^2}\sim0,\quad \bar\eta\simeq\frac{\e^2}{4(|x|+t)^2}\sim0
\]
in the limits of $\e\to0$. In terms of the expansion \eqref{conformal-block-at-0} of the conformal blocks at zero, we arrive at the early time behavior of the four-point function $G^{21}_{34}$,
\[
G^{21}_{34}(\eta,\bar\eta)=\sum_{p}\left(C^p_{12}C^p_{34}\left(\frac{\e^2}{4|x|^2-4t^2}\right)^{2h_p-2h_3-2h_4}\big(1+O(\e)\big)\right).\label{becom1}
\]
On the other hand, at early times, the leading order of the prefactor of $G^{21}_{34}$ in eqn. \eqref{4pt-Sigma2->Sigma1} in $\e\to0$ limit reads
\[
\left(\bigg|\frac{2z_1z_{12}z_{14}}{z_{24}}\bigg|^{-2h_1}\prod_{i=2}^4\bigg|\frac{2z_iz^2_{1i}z_{24}}{z_{12}z_{14}}\bigg|^{-2h_i}\right)=\frac{2^{-2(h_1+h_2+h_3+h_4)}\e^{-2h_1-2h_2+2h_3+2h_4}}{(4|x|^2-4t^2)^{2h_3+2h_4}}\big(1+O(\e)\big).\label{becom2}
\]
Combining \eqref{becom1} with \eqref{becom2}, the four-point function on $\Sigma_2$ at early times is given by
\[
&(2\e)^{2(h_1+h_2+h_3+h_4)}\la\mo_1(w_1,\bar w_1)\mo^\dagger_2(w_2,\bar w_2)\mo_3(w_3,\bar w_3)\mo_4^\dagger(w_4,\bar w_4)\ra_{\Sigma_2}\nn\\
=&\sum_p\left( C^p_{12}C^p_{34}\left(\frac{\e^2}{4|x|^2-4t^2}\right)^{2h_p}\big(1+O(\e)\big)\right)\quad(0<t<|x|).
\]
It can be observed that the above equation is non-zero if and only if there exists an identity block in $G_{34}^{21}$. In this scenario, we have $\mo_3=\mo_4$ and $\mo_1=\mo_2$. Consequently, we conclude that the four-point functions on $\Sigma_2$ at early time
\[
(2\e)^{2(h_i+h_j+h_k+h_l)}\la\mo_i(w_1,\bar w_1)\mo^\dagger_j(w_2,\bar w_2)\mo_k(w_3,\bar w_3)\mo_l^\dagger(w_4,\bar w_4)\ra_{\Sigma_2}=\d_{ij}\d_{kl}.\label{2delta}
\]
Substituting \eqref{2delta} into \eqref{2eesec3}, we find that the 2nd REE at early times is zero,
\[
\Delta S^{(2)}\big(\rho_A^{\psi}(t)\big)=-\log\sum_{ijkl\in\S}C_iC_j^*C_kC_l^*\d_{ij}\d_{kl}=-2\log\sum_{i\in\S}|C_i|^2=0
\quad(0<t<|x|).
\]

The late-time analysis closely parallels that of section \ref{section: replica method}. It's essential to note that as we are discussing EE rather than PE (i.e., we have $\tilde x=x$), the cross-ratio will experience a discontinuity at $t=|x|$,
\[
\eta\simeq1-\frac{\e^2}{4(t-|x|)^2}\sim1,\quad \bar\eta\simeq\frac{\e^2}{4(t+|x|)^2}\sim0,\quad(t>|x|).
\]
This results in
\[
G^{21}_{34}(\eta,\bar\eta)=\sum_{p,q}\left(C_{12}^{p}C_{34}^{p}
F_{pq}\left[\begin{smallmatrix}2,1\\3,4\end{smallmatrix}\right] \frac{(\e/2)^{2(h_p+h_q-h_2-2h_3-h_4)}}{(t-|x|)^{2(h_q-h_2-h_3)}(t+|x|)^{2(h_p-h_3-h_4)}}\big(1+O(\e)\big)\right)
\]
for $t>|x|$. On the other hand, the leading order of the prefactor at late times reads
\[
&\left(\bigg|\frac{2z_1z_{12}z_{14}}{z_{24}}\bigg|^{-2h_1}\prod_{i=2}^4\bigg|\frac{2z_iz^2_{1i}z_{24}}{z_{12}z_{14}}\bigg|^{-2h_i}\right)\nn\\
=&\e^{2h_3-2h_1}4^{-h_1-2h_2-3h_3-2h_4}(t-|x|)^{-2(h_2+h_3)}(t+|x|)^{-2(h_3+h_4)}\big(1+O(\e)\big).
\]
Combining the above two equations, we find that
\[
&(2\e)^{2(h_1+h_2+h_3+h_4)}\la\mo_1(w_1,\bar w_1)\mo^\dagger_2(w_2,\bar w_2)\mo_3(w_3,\bar w_3)\mo_4^\dagger(w_4,\bar w_4)\ra_{\Sigma_2}\nn\\
=&\sum_{pq}\left( C^p_{12}C^p_{34}F_{pq}\left[\begin{smallmatrix}2,1\\3,4\end{smallmatrix}\right]\frac{\e^{2(h_p+h_q)}}{(t-|x|)^{2h_q}(t+|x|)^{2h_p}}\big(1+O(\e)\big)\right)\quad(t>|x|).\label{123q}
\]
Similarly, we observe that eqn. \eqref{123q} is non-zero if and only if both vacuum blocks exist in $G_{34}^{21}$ and $G^{41}_{32}$. In this scenario, we have  $\mo_1=\mo_2=\mo_3=\mo_4$. We thus conclude that at late times
\[
(2\e)^{2(h_i+h_j+h_k+h_l)}\la\mo_i(w_1,\bar w_1)\mo^\dagger_j(w_2,\bar w_2)\mo_k(w_3,\bar w_3)\mo_l^\dagger(w_4,\bar w_4)\ra_{\Sigma_2}=
\begin{cases}
d_i^{-1},&\quad i=j=k=l,\\
0,&\text{otherwise}.
\end{cases},\label{22delta}
\]
where we use the formula  $F_{00}\left[\begin{smallmatrix}\mo,\mo\\\mo,\mo\end{smallmatrix}\right]=d^{-1}_{\mo}$
 in RCFTs \cite{Moore:1988ss}. Substituting \eqref{22delta} into \eqref{2eesec3}, one obtains the late-time value which was previously encountered in section \ref{section2.3}.

We have demonstrated the 2nd REE evolution of the linear combination operator \eqref{normalized-linear-combination-operator} exhibiting a step-like behavior. The results can be straightforwardly extended to a general $n$ based on the methodology outlined in section \ref{section: replica method}. For the general $n$-th REE evolution, the $2n$-point functions on $\Sigma_n$ behave like
\[
(2\e)^{2\sum^{2n}_{\a=1}h_{i_\a}}\Big\la\prod^{n}_{\a=1}\mo_{i_{2\a-1}}(w_{2\a-1},\bar w_{2\a-1})\mo^\dagger_{i_{2\a}}(w_{2\a},\bar w_{2\a})\Big\ra_{\Sigma_2}=
\begin{cases}
\prod_{\a=1}^{n}\d_{i_{2\a-1}i_{2\a}},&\quad 0<t<|x|,\\
d^{1-n}_{i_1}\prod^{2n}_{\a=2}\d_{i_1i_{\a}},&\quad t>|x|.\label{ndelta}
\end{cases}
\]
Consequently, the complete time evolution of the $n$-th REE is given by
\[\Delta S^{(n)}\big(\rho_A^{\psi}(t)\big)=
\begin{cases}
0,&\quad 0<t<|x|,\\
\frac{1}{1-n}\log\sum_{k\in\S}d^{1-n}_{k}|C_k|^{2n},&\quad t>|x|.
\end{cases}\label{tosub2}
\]
By analytically continuing $n$ to 1, one obtains the complete time evolution of EE,
\[\Delta S\big(\rho_A^{\psi}(t)\big)=
\begin{cases}
0,&\quad 0<t<|x|,\\
H(p_k)+\sum_{k\in\S}p_k\log d_k,&\quad t>|x|,
\end{cases}\label{SA}
\]
where $p_k\equiv|C_k|^2$ and $H(p_k)$ the classical Shannon entropy of $\{p_k\}$.  Eqn. \eqref{SA} indicates that the classical picture of quasiparticle propagation remains applicable for the linear combination states \eqref{psiyiban} in RCFTs. It should be noted that while the evolution of EE and REE exhibits a step-like behavior, in the case of PE and PREE, the results generally do not follow a step-like pattern \cite{Guo:2022sfl}. That is, the late-time formulae for PE \eqref{PEE_final} and PREE \eqref{nth-PREE-explict-form} hold only under large-$t$ limit.
\vskip.4em
\subsection{Block diagonalization of the reduced density matrix at late times}
What we aim to elaborate on next is that eqn. \eqref{SA} suggests that after $t$ exceeds $|x|$, the reduced density matrix $\rho_A^{\psi}$ will be block diagonal\footnote{We would like to thank Wu-zhong Guo for valuable discussions regarding this section.},
\[
\rho^{\psi}_A(t)=\underset{k\in\S}{\oplus}\rho^{\mo_k}_A(t),\quad(t>|x|),\label{block dia}
\]
where $\rho^{\mo_k}_A(t)$ represents the reduced density matrix \eqref{rhoak} generated by the primary operator $\mo_k$ within the set $\S$. To see this, let us first define a new reduced density matrix of $A$ utilizing $\rho_A^{\mo_k}$ \cite{Bhattacharyya:2019ifi},
\[
\sigma^{\psi}_A(t):=\sum_{k\in\S}|C_k|^2\rho_A^{\mo_k}(t),\quad \tr_A[\sigma^\psi_A(t)]=1.
\]
Despite the apparent difference between $\rho^{\psi}_A$ and $\sigma_A^{\psi}$ by definition,
\[
\rho^{\psi}_A-\sigma^{\psi}_A=&\sum_{i,j\in\S;i\neq j}C_iC_j^*\T^{\mo_i|\mo_j},\nn\\
\Bigg(\T^{\mo_i|\mo_j}(t):=&|\mo_i(x,t)\ra\la\mo_j(x,t)|\equiv\frac{e^{-iHt}\mo_i(x,-\e)|\Omega\ra\la\Omega|\mo_j^\dagger(x,\e)e^{iHt}}{\sqrt{\la\mo^\dagger_i(x,\e)\mo_i(x,-\e)\ra\la\mo^\dagger_j(x,\e)\mo_j(x,-\e)\ra}}\Bigg),
\]
we demonstrate that indeed $\rho^{\psi}_A$ and $\sigma^{\psi}_A$ are equivalent throughout the entire time evolution. This can be achieved by computing the relative entropy between $\rho^{\psi}_A$ and $\sigma^{\psi}_A$. We know that the relative entropy between two reduced density matrices is 0 if and only if these two reduced density matrices are identical. In practice, one usually  compute the R\'enyi relative entropy\footnote{Unlike the relative entropy, R\'enyi relative entropy is not non-negative.} \cite{Lashkari:2015dia}
\[
S^{(n)}\big(\rho^{\psi}_A||\sigma_A^{\psi}\big)=\frac{1}{n-1}\Big(\log\tr_A[(\rho^{\psi}_A)^n]-\log\tr_A[\rho_A^{\psi}(\sigma_A^{\psi})^{n-1}]\Big)\label{nthRREE}
\]
first and then take the limit of $n\to1$ to obtain the relative entropy,
\[
S\big(\rho^{\psi}_A||\sigma_A^{\psi}\big)=\lim\limits_{n\to1}S^{(n)}\big(\rho^{\psi}_A||\sigma_A^{\psi}\big)=\tr_A[\rho_A^{\psi}\log\rho_A^{\psi}]-\tr_A[\rho^{\psi}_A\log\sigma_A^{\psi}].
\]
Since the first part of the $n$-th R\'enyi relative entropy \eqref{nthRREE} is nothing but the $n$-th REE of $\rho^{\psi}_A$, we only need to focus on its second part, which amounts to the following sum of $2n$-point functions on $\Sigma_n$,
\[
\tr_A[\rho_A^{\psi}(\sigma_A^{\psi})^{n-1}]=&\frac{Z_n}{Z_1^n}\sum_{\{k_\a\}\in\S}C_{k_0}C^*_{k_1}\prod^{n}_{\a=2}|C_{k_\a}|^2(2\e)^{2(h_{k_0}+h_{k_1}+2\sum^{n}_{\a=2}h_{k_\a })}\nn\\
&\times\Big\la\mo_{k_0}(w_1,\bar w_1)\mo^\dagger_{k_1}(w_2,\bar w_2)\prod^{n}_{\a=2}\mo_{k_{\a}}(w_{2\a-1},\bar w_{2\a-1})\mo_{k_\a}^\dagger(w_{2\a},\bar w_{2\a})\Big\ra_{\Sigma_n}.
\]
Leveraging eqn. \eqref{ndelta}, we can  figure out the above sum and determine the evolution behavior of $\tr_A[\rho_A^{\psi}(\sigma_A^{\psi})^{n-1}]$\footnote{Recall that $Z_n$ is the partition function of the Riemann surface $\Sigma_n$.},
\[
\tr_A[\rho_A^{\psi}(\sigma_A^{\psi})^{n-1}]=
\begin{cases}
\frac{Z_n}{Z_1^n},&\quad 0<t<|x|,\\
\frac{Z_n}{Z_1^n}\sum_{k\in\S}d_k^{1-n}|C_k|^{2n},&\quad t>|x|.\label{tosub1}
\end{cases}
\]
Substituting eqns \eqref{tosub1} and \eqref{tosub2} into eqn. \eqref{nthRREE}, it becomes evident that the R\'enyi relative entropy of any order remains zero throughout time evolution. Utilizing the analytic continuation of $n$, the relative entropy between $\rho_A^{\psi}$ and $\sigma_A^{\psi}$ stays zero consistently,
\[
 S\big(\rho_A^{\psi}(t)||\sigma_A^{\psi}(t)\big)=\lim\limits_{n\to1}S^{(n)}\big(\rho_A^{\psi}(t)||\sigma_A^{\psi}(t)\big)=0,\quad(t>0).
\]
Therefore, we conclude that $\rho_A^{\psi}(t)=\sigma_A^{\psi}(t)=\sum_{k\in\S}p_k\rho_A^{\mo_k}(t)$ throughout the evolution. Next, we demonstrate that the reduced density matrices corresponding to different primary operators are orthogonal to each other at late times. The crux of the matter is that we can find
\[
\tr_A\Big[\sqrt{\rho_A^{\mo_l}(t)}\rho_A^{\mo_k}(t)\sqrt{\rho_A^{\mo_l}(t)}\Big]=\tr_A[\rho_A^{\mo_k}(t)\rho_A^{\mo_l}(t)]=0, \quad (t>|x|,~k\neq l)
\]
through eqn. \eqref{2delta}. Since $\sqrt{\rho_A^{\mo_l}(t)}\rho_A^{\mo_k}(t)\sqrt{\rho_A^{\mo_l}(t)}$ is a semi-positive definite operator, the above equation being valid implies that $\sqrt{\rho_A^{\mo_l}(t)}\rho_A^{\mo_k}(t)\sqrt{\rho_A^{\mo_l}(t)}$ has a spectrum consisting solely of zeros. On the other hand, by the Jacobson Lemma, the operators $AB$ and $BA$ share the same nonzero spectrum. Hence, $\rho_A^{\mo_k}(t)\rho_A^{\mo_l}(t)$ also possesses a spectrum solely composed of zeros. {Moreover, owing to the semi-positivity of $\rho_A^{\mo_k}(t)$ and $\rho_A^{\mo_l}(t)$, $\rho_A^{\mo_k}(t)\rho_A^{\mo_l}(t)$ is diagonalizable(see appendix \ref{appendixA} for more details), indicating that $\rho_A^{\mo_k}(t)\rho_A^{\mo_l}(t)$ is zero.} Similarly, $\rho^{\mo_l}_A(t)\rho^{\mo_k}_A(t)$ is also a zero operator, so $\rho^{\mo_k}_A(t)$ and $\rho_A^{\mo_l}(t)$ are commutative.

When two matrices $\rho_i$ and $\rho_j$ are commutative, there exist a unitary matrix denoted as $U$ such that
\[
U\rho_i U^{\dagger}=\Lambda_i,\quad U\rho_j U^{\dagger}=\Lambda_j,\label{eq3.2.1}
\]
where $\Lambda_i$ and $\Lambda_j$ are diagonal matrices with non-negative eigenvalues.  In this scenario, $\rho_i\rho_j=0$ is equivalent to $\Lambda_i \Lambda_j =0$. The latter implies that the nonzero eigenvalues of $\Lambda_i$ and the nonzero eigenvalues of $\Lambda_j$ are distributed at different locations along the diagonal. Consequently, $\rho_i+\rho_j$ will be a block diagonal matrix. Therefore, we conclude that $\rho_A^{\psi}(t)=\sum_{k\in\S}p_k\rho_A^{\mo_k}(t)$ will be a block diagonal matrix \eqref{block dia} after $t>|x|$.

\section{Conclusion and prospect}\label{sec4}
This paper investigates the late-time behavior of REE and PRE in local operator quenches of RCFT, with a particular emphasis on quenches involving linear combination operators.
For a state $|\psi\rangle$ in the form of \eqref{simplified-linear-combination-operator}, the REE captures only the information of the heaviest primary \eqref{REE-for-heavy-operators}. In contrast, when primary operators with distinct conformal weights are included, the PREE of the reduced transition matrix $\T^{\psi|\tpsi}_A$ preserves the information of all the primary operators \eqref{nth-PREE-explict-form}. The maximal value attainable by $S^{(n)}(\T_A^{\psi|\tpsi})$ but unattainable by $S^{(n)}(\rho_A^{\psi})$ is the logarithm of the sum of the quantum dimensions of all primary operators,
\[
\max_{\{C_k\}}\left\{\lim_{t\to\infty}\Delta S^{(n)}(\mathcal{T}_A^{\psi|\tilde\psi}(t))\right\}=\log\left(\sum_{k\in\S}d_k\right).
\]
This phenomenon, also known as ``pseudo-entropy amplification", occurs when incorporating operators with different conformal dimensions within the linear combination operator $\mathcal{O}$ in \eqref{simplified-linear-combination-operator}.

Furthermore, we utilized EE to explore the relationships among quantum dimension, OPE, and fusion coefficients. The computation of EE from both sides of eqn  \eqref{OPE-state} should yield identical results, leading to the relationship between OPE coefficients and fusion coefficients
\[
\sqrt{\frac{N_{12}^kd_k}{N_{12}^ld_l}}=\frac{C_{k12}}{C_{l12}}\lim_{\e\to0}\sqrt{\frac{|\mathcal{N}_{k12}(\e)|}{|\mathcal{N}_{l12}(\e)|}},\quad N^{k}_{12}\in\{0,1\}.
\]
%It should be noted that this equation involves the consideration of an infinite number of secondary fields.

%{\color{} Finally, we discuss the evolution of the EE. As mentioned in \cite{}, the state \eqref{} is unnatural }

Finally, we discuss that to recover information about lighter operators from EE in general cases, one should consider the regularized linear combination operator \eqref{normalized-linear-combination-operator} instead of \eqref{simplified-linear-combination-operator}, or equivalently, from the perspective of states,
%In this definition, each primary in the sum is normalized by its two-point function. While this definition may appear unconventional since $\epsilon$ has to approach zero, leading to divergent coefficients, it's perhaps more natural from the perspective of linear combinations of locally primary excited states
\[
|\psi\rangle=\frac{1}{\sqrt{\mathcal{N}(\epsilon)}}\mo(x;\epsilon)|\Omega\rangle=\sum_{k\in\S}C_k|\mo_k(x)\ra.
\]
It is in combining states within the Hilbert space that we truly perform linear combinations, highlighting the nontriviality of defining the linear combination operator from the perspective of operators.
The full-time evolution of EE for the regularized linear combination operator is determined to be
\[
\Delta S\big(\rho_A^{\psi}(t)\big)=
\begin{cases}
0,&\quad 0<t<|x|,\\
H(p_k)+\sum_{k\in\S}p_k\log d_k,&\quad t>|x|,
\end{cases}
\]
indicating that the classical picture of quasiparticles propagation remains applicable for the regularized linear combination states in RCFTs. The quasiparticle picture also indicate that the reduced density matrix $\rho_A^{\psi}$ will be block diagonal after $t$ surpasses  $|x|$,
\[
\rho^{\psi}_A(t)=\underset{k\in\S}{\oplus}\rho^{\mo_k}_A(t),\quad(t>|x|),
\]
with each block containing information about one conformal family.

{Since we are currently focusing on RCFTs, a natural question for future research is to extend the relevant investigations to Liouville CFT, holographic CFTs, $T\bar T$-deformed CFTs, etc.} When investigating the relationship between quantum dimension, OPE coefficients, and fusion coefficients, we must deal with a summation involving OPE coefficients $\beta$ and coefficients of two point functions, which is extremely challenging at the technical level. The block diagonalization of the reduced density matrix for the linear combination operator, akin to symmetry resolved entanglement \cite{Goldstein:2017bua,Zhao:2020qmn,Capizzi:2020jed,Weisenberger:2021eby,Kusuki:2023bsp}, may reveal hidden symmetries in the local linear combination operator quenches. We would like to leave them to future work.
%The late-time $n$-th REE and EE for the operator $\mo(x,\epsilon)$ remain as given in eqn. $\eqref{nthREE-linear-combination-operator}$ and eqn. $\eqref{EE-of-rhoA}$. Therefore, the maximum and the corresponding coefficients change.
\subsection*{Acknowledgements}
We would like to thank Tadashi Takayanagi, Wu-zhong Guo, Yuya Kusuki, Yuan Sun, Hao Ouyang, Hong-An Zeng, and Yang Liu for their valuable discussions related to this work. We are also grateful to all the organizers of the "Quantum Information, Quantum Matter and Quantum Gravity" workshop (YITP-T-23-01) held at YITP, Kyoto University, where a part of this work was done.  SH would like to appreciate the financial support from Jilin University, Max Planck Partner Group, and the Natural Science Foundation of China Grants (No.12075101, No.12235016). L.Z and Z.Z are supported by the Science and Technology Development Plan Project of Jilin Province, China (No. 20240101326JC).
\appendix
\section{The product of two semi-positive matrices is diagonalizable}\label{appendixA}
{Suppose $A$ and $B$ are two semi-positive matrices, so there exist a invertible matrix $P$, such that $B=P^\dagger DP$, where
\[
D=\begin{pmatrix}
I_k & 0 \\
0 & 0
\end{pmatrix},
\]
 where $I_k$ denotes a $k$ by $k$ identity matrix.
Since $AB\sim PAP^\dagger(P^\dagger)^{-1}BP^{-1}=PAP^\dagger D\equiv CD$, to prove the product of two semi-positive matrices is diagonalizable, we need to show matrix $CD$ is diagonalizable.
Let matrix $C$ be equal to
\[
\begin{pmatrix}
C_1 & C_2^\dagger \\
C_2 & C_3
\end{pmatrix},
\]
and by the definition of $C$, it is obviously a semi-positive matrix which can be reformulated as
\[
C=M^\dagger\begin{pmatrix}
I_{k'} & 0 \\
0 & 0
\end{pmatrix}M,
\]
where $M$ is a invertible matrix.

Suppose $x$ is the solution of equation $C_1 x=0$, then we have
\[
\begin{pmatrix}
    x^\dagger & 0
\end{pmatrix}
\begin{pmatrix}
    C_1 & C_2^\dagger \\
C_2 & C_3
\end{pmatrix}
\begin{pmatrix}
    x\\
    0
\end{pmatrix}=\begin{pmatrix}
    x^\dagger & 0
\end{pmatrix}M^\dagger
\begin{pmatrix}
    I_{k'} & 0 \\
0 & 0
\end{pmatrix}M
\begin{pmatrix}
    x\\
    0
\end{pmatrix}=0.\label{a.1}
\]
If we denote $y=M\begin{pmatrix}
    x\\
    0
\end{pmatrix}$, \eqref{a.1} can be rewritten as
\[
y^\dagger\begin{pmatrix}
    I_{k'} & 0 \\
0 & 0
\end{pmatrix}y=0,
\]
giving rise to the fact that $y=\begin{pmatrix}
    0\\
    *
\end{pmatrix}$ which leads to $y^\dagger\begin{pmatrix}
    I_{k'} & 0 \\
0 & 0
\end{pmatrix}M=0$, i.e.
\[
\begin{pmatrix}
    x^\dagger & 0
\end{pmatrix}
\begin{pmatrix}
    C_1 & C_2' \\
C_2 & C_3
\end{pmatrix}=0\Rightarrow\begin{pmatrix}
    C_1\\
    C_2
\end{pmatrix}x=0.
\]
Therefore, $C_1 x=0 \iff\begin{pmatrix}
    C_1\\
    C_2
\end{pmatrix}x=0$, i.e. $rank(C_1)=rank\begin{pmatrix}
    C_1\\
    C_2
\end{pmatrix}$, indicating that there exist a matrix $Q$ such that $C_2=Q C_1$. After multiplying the matrix CD on the left by a matrix $X$ and on the right by the inverse of $X$, where $X=\begin{pmatrix}
    I & 0\\
    -Q&I
\end{pmatrix}$, $CD$ becomes
\[
\begin{pmatrix}
    I & 0\\
    -Q&I
\end{pmatrix}CD
\begin{pmatrix}
    I & 0\\
    Q&I
\end{pmatrix}=\begin{pmatrix}
    I & 0\\
    -Q&I
\end{pmatrix}\begin{pmatrix}
    C_1&0\\
    C_2&0
\end{pmatrix}\begin{pmatrix}
    I & 0\\
    Q&I
\end{pmatrix}=\begin{pmatrix}
    C_1 & 0\\
    0&0
\end{pmatrix}.
\]
Since $C$ is a semi-positive matrix, $C_1$ can be diagonalized, i.e. $CD$ is diagonalizable which leads to the fact that the product of two semi-positive matrices is diagonalizable.}
\bibliographystyle{JHEP}
\bibliography{PEA-bib}{}
\end{document}